\documentclass[journal=jacsat,manuscript=article]{achemso}

\usepackage[LGRgreek]{mathastext}
\usepackage{url}
\usepackage[version=3]{mhchem} 
\usepackage[dvipsnames]{xcolor}
\usepackage{lmodern,adjustbox,booktabs,multirow}

\usepackage[english]{babel}
\author{Jaclyn R. Lunger}
\affiliation{Department of Materials Science and Engineering, Massachusetts Institute of Technology, Cambridge, Massachusetts 02139 United States}
\author{Jessica Karaguesian}
\affiliation{Center for Computational Science and Engineering, Massachusetts Institute of Technology, Cambridge, Massachusetts 02139 United States}
\author{Hoje Chun}
\affiliation{Department of Chemical and Biomolecular Engineering, Yonsei University, Seoul 03722 Republic of Korea}
\author{Jiayu Peng}
\affiliation{Department of Materials Science and Engineering, Massachusetts Institute of Technology, Cambridge, Massachusetts 02139 United States}
\author{Yitong Tseo}
\affiliation{Department of Computational and Systems Biology, Massachusetts Institute of Technology, Cambridge, Massachusetts 02139 United States}
\author{Chung Hsuan Shan}
\affiliation{Department of Materials Science and Engineering, University of Toronto, Toronto ON M5S Canada}
\author{Byungchan Han}
\affiliation{Department of Chemical and Biomolecular Engineering, Yonsei University, Seoul 03722 Republic of Korea}
\author{Yang Shao-Horn}
\affiliation{Department of Materials Science and Engineering, Massachusetts Institute of Technology, Cambridge, Massachusetts 02139 United States}
\altaffiliation{Department of Mechanical Engineering, Massachusetts Institute of Technology, Cambridge, Massachusetts 02139 United States}
\alsoaffiliation{Research Laboratory of Electronics, Massachusetts Institute of Technology, Cambridge, Massachusetts 02139 United States}
\email{shaohorn@mit.edu}
\author{Rafael G\'{o}mez-Bombarelli}
\affiliation{Department of Materials Science and Engineering, Massachusetts Institute of Technology, Cambridge, Massachusetts 02139 United States}
\email{rafagb@mit.edu}

\title[Atom-by-atom design of metal oxide catalysts]
  {Atom-by-atom design of metal oxide catalysts for the oxygen evolution reaction with machine learning}

\abbreviations{IR,NMR,UV}

\begin{document}



\section{Summary}

Green hydrogen production is crucial for a sustainable future, but current catalysts for the oxygen evolution reaction (OER) suffer from slow kinetics, despite many efforts to produce optimal designs, particularly through the calculation of descriptors for activity. In this study, we develop a dataset of density functional theory calculations of bulk and surface perovskite oxides, and adsorption energies of OER intermediates, which includes compositions up to quaternary and facets up to (555). We demonstrate that per-site properties of perovskite oxides such as Bader charge or band center can be tuned through element substitution and faceting, and develop a machine learning model that accurately predicts these properties directly from the local chemical environment. We leverage these per-site properties to identify promising perovskites with high theoretical OER activity. The identified design principles and promising new materials provide a roadmap for closing the gap between current artificial catalysts and biological enzymes.

\section{Keywords}

atomic design, catalysis, enzymes, heterogeneous, faceting, substitution, graph neural, perovskite, DFT, machine learning


\section{Context and Scale}

The demand for sustainable energy sources has never been more pressing, and developing innovative, efficient, and eco-friendly energy technologies is essential. The oxygen evolution reaction (OER) holds great promise, as it enables the electrolytic production of green chemicals and hydrogen gas, with oxygen as the sole byproduct. However, OER suffers from a lack of efficient artificial catalysts and from slow kinetics. In contrast, enzymes such as photosystem II exhibit much higher efficiency, as a result of millions of years of biological evolution over a complex design space. Here, we employ computation to navigate the landscape of metal oxide catalysts for OER, looking for optimal designs to rival the activity of enzymes. This work uses a systematic approach beyond traditional trial-and-error approaches, by combining machine learning, site-level descriptors, and high-throughput virtual screening using density functional theory to identify atom-by-atom design principles that enhance the efficient production of hydrogen.

\section{Introduction}

Atomic-level design of highly active catalysts that rival the efficiency of natural enzymes is one of the ultimate goals of chemistry and materials science. Whereas enzymes have been fine-tuned residue-by-residue through biological evolution for millions of years, human-designed homo- and hetero-geneous catalysts to date remain far from optimal. For example, for the oxygen evolution reaction (OER), one of the most important reactions in both photosynthesis and renewable chemical and fuel production, the turnover frequency in photosystem II is at least two or three orders of magnitude higher than state-of-the-art heterogeneous catalysts\cite{Hong2015, Dau2010, Cox2014}. Moreover, enzymatic catalysis also demonstrates the significance of site-dependent reactivity, as reactions occur within highly specialized pockets and tailored chemical environments\cite{Vogt2022, Cox2014}.
 
Efforts to design heterogeneous catalysts with activities rivaling enzymes will require breaking scaling relations, which dictate that binding energies of reaction intermediates are linearly dependent and thus limit the minimum theoretical overpotential to $\sim{0.37}$ V\cite{Norskov2014,ferri2023approaching}. Of the various strategies for breaking scaling, facet engineering\cite{Rao2020} and element substitution\cite{Halck2014} remain some of the simplest and most widely used. Multicomponent metal oxide surfaces, which take advantage of both of these strategies, thus offer a unique opportunity to move beyond these limitations, due to their vast design space comprising billions of possible local atomic structures and active site environments. However, exploration of such a vast chemical space to design optimal artificial active sites will require atom-by-atom design beyond current approaches utilizing trial-and-error coupled with human intuition\cite{Peng2022}.

One effective strategy to go beyond Edisonian design of complex oxide catalysts is to use a descriptor-centered approach, which captures the quantitative relationship between structure, energetics, and function\cite{Peng2022}. Descriptors are especially powerful as they can be calculated \textit{in silico} with low cost, such that materials with desired properties can be predicted before they are even synthesized. For instance, catalytic activity has been shown to correlate with d-band center for metals\cite{Hammer1995}, and oxygen 2p-band center for metal oxides\cite{Grimaud2013,Jacobs2018}. Moreover, $e_g$ orbital filling\cite{Suntivich2011}, spin state\cite{Biz2021,Vennelakanti2022} and magnetic ordering\cite{Biz2021} have also been shown to have a profound effect on catalytic activity. While the use of such descriptors in catalyst designs has already led to successful catalyst discoveries\cite{Grimaud2013,Grimaud2017, Kuznetsov2020,Yuan2022}, most of these previous efforts have mainly relied on bulk descriptors, which assumes that surface electronic structure is correlated with bulk electronic structure characteristics\cite{CalleVallejo2013}.

Despite their effectiveness, average bulk descriptors alone cannot capture local effects from substitution or surface effects, limiting the space of atom-by-atom design for surface reactivity\cite{Dickens2017}. Such limitation is particularly severe for complex multimetallic oxide surfaces\cite{Choubisa2023} due to the diverse and near-continuous distributions of unique surface sites. Moreover, it has been recently shown that for RuO$_2$, a state-of-the-art catalyst for the oxygen evolution reaction (OER), the catalytic activity is highly dependent on the local environment of coordinatively unsaturated surface oxygen\cite{Rao2020,Rao2017}. Specifically, CUS-oxygen as the active sites on (100) is shown to have an order of magnitude higher specific OER activity that of (110)\cite{Rao2020}. Design of heterogeneous catalysts with activity rivaling that of enzymes thus requires the understanding and efficient prediction of site-specific activity. 

Budding efforts to go beyond bulk descriptors and to link local, site-dependent descriptors to catalytic activity include the report that the site-dependent 2p-band center of adsorbed oxygen correlates strongly with OER-intermediate binding energies\cite{Dickens2019,Hwang2021}. Nevertheless, this proposed descriptor requires the same DFT calculation as calculating the binding energy itself so it does not result in a significant computational advantage. Site-dependent structural descriptors, such as generalized coordination number, are cheaper to calculate since they only require the pristine slab and have been shown to correlate with intermediate binding energies\cite{CalleVallejo2015, Ruck2018}. While coordination number is able to capture local structure-property trends, it alone cannot be generalized across different local chemistry and coordination. Recent efforts to overcome these limitations have included encoding elemental as well as coordination information\cite{Batchelor2019, Tran2018}, often relying on hand-crafted features that are difficult to generalize and extend to new materials classes\cite{Fernandez2017,Li2017}. Machine learning approaches have revolutionized computational materials science\cite{Axelrod2022,Kitchin2018} given their ability to learn very complex non-linear functions, efficiently process structural information without hand-tuning\cite{Xie2018,Damewood2023}, and combine this information with other data fidelities\cite{Greenman2022} or physics-based priors, typically through graph-convolutional neural networks. Some recent  works have leveraged these tools to predict OER-intermediate binding energies directly from structure\cite{Tran2022}, circumventing the need for hand-crafted features, but the models to date have not taken advantage of site-dependent descriptors beyond local structure\cite{Back2019}.

In this work, we develop a graph-based neural network model that accurately learns site-dependent descriptors from structure and leverages these descriptors to predict OER intermediate binding energies automatically from structure, without the need for hand-crafted features. We consider density functional theory (DFT)-computed site-specific descriptors based on local electronic structure (Bader charges\cite{Hwang2017,Rawal2019}, site-projected O 2p-band\cite{Dickens2019,Hwang2021} and metal d-band centers\cite{Hammer1995}), local magnetic moments\cite{Zhong2021} with previous datasets\cite{Castelli2012, Jacobs2018, Emery2016, Jain2013, Tran2022}, we develop and curate, to the best of our knowledge, the largest and most diverse dataset of OER energetics on perovskite surfaces to date. We show that per-site descriptors across this dataset vary widely with composition and surface coordination, and can be predicted purely from structure. We demonstrate per-site properties that are both independent of one another and linearly correlated to OER binding energies, such that they can be leveraged independently to tune OER energetics. We compare the OER energetics across this dataset with previous efforts on metals\cite{Kulkarni2018}, rutile RuO$_2$\cite{Rao2017,Rao2020} and IrO$_2$\cite{Gauthier2017}, metal hydroxide-organic frameworks (MHOFs)\cite{Yuan2022}, single atom catalysts\cite{Deng2019}, and enzymes\cite{Siegbahn2013}, and show that by tuning per-site properties of the active site and adsorbed species it is theoretically possible to extend beyond the scaling relations and bridge the gap between current state-of-the-art heterogeneous catalysts and enzymes. Finally, we list several of the most promising materials identified through this work.

\section{Results and discussion}

\begin{figure}
    \begin{center}
    \includegraphics[width=\textwidth]{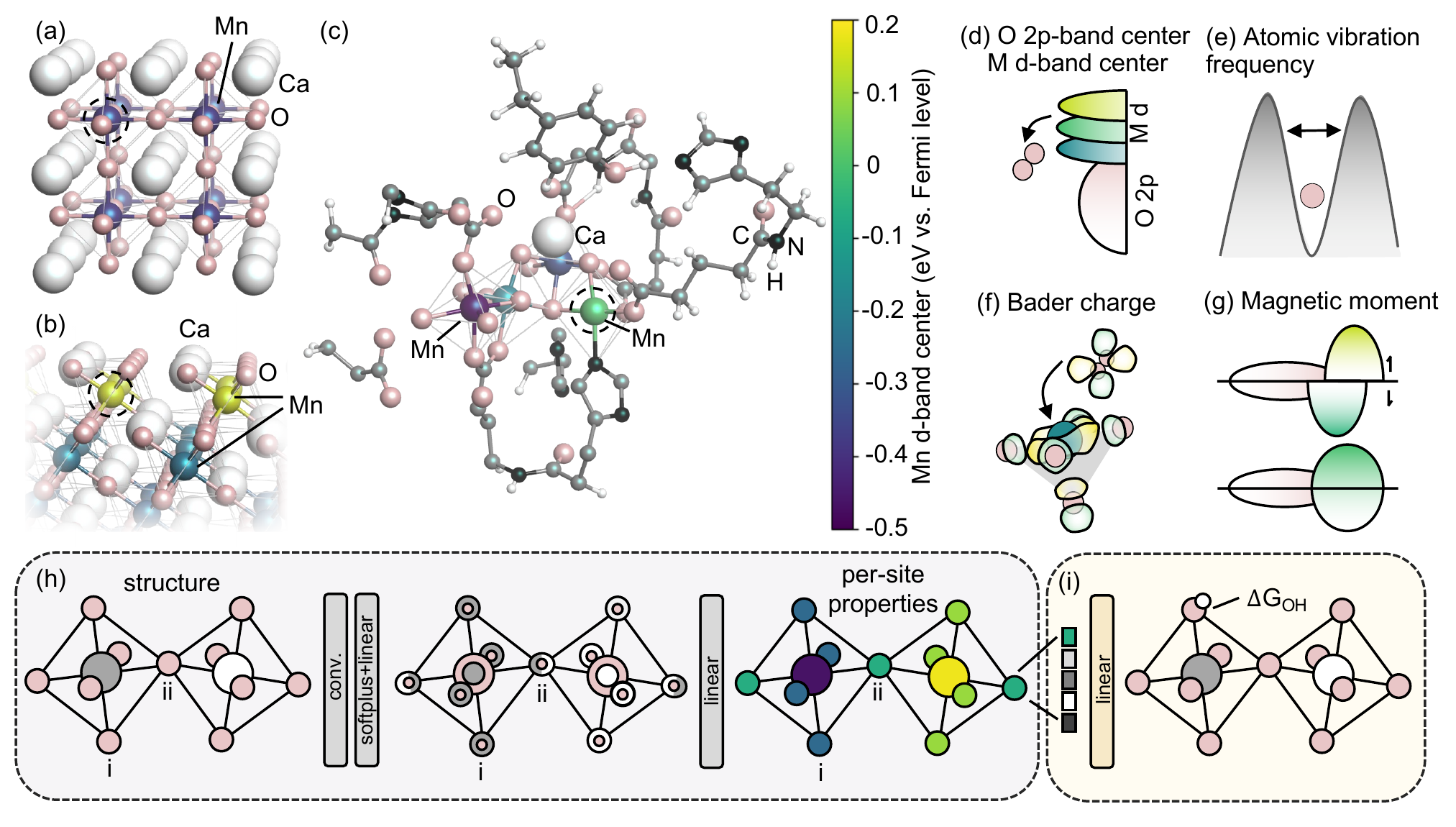}
    \end{center}
     \caption{Per-site properties. Structures for (a) bulk CaMnO$_3$, (b) CaMnO$_3$ (111) surface, and (c) PSII with Mn atoms colored by d-band center (eV vs. Fermi level) and the active site circled with a black dashed line. The structure for PSII is taken from Lohmiller et al., and a single self consistent field cycle is done to extract band centers\cite{Lohmiller2014}. (d-g) Site-level material descriptors relevant for atom-by-atom catalyst designs, including O 2p-band center, M d-band center, atomic vibration frequency, Bader charge, and magnetic moment. Schematic of the model architecture for (h) predicting and leveraging per-site properties to (i) ultimately learn binding energies. The explicit mapping between local environment and site property is made evident by following oxygen atom i and ii, which have different local environments leading to different convolved representations and site properties.}
     \label{architecture}
\end{figure}

The local environment around the active site strongly influences the materials properties of that site. For example, the Mn atoms in bulk CaMnO$_3$ (Figure \ref{architecture}a), on the CaMnO$_3$ $(111)$ surface (Figure \ref{architecture}b), and inside photosystem II (PSII) have different local environments and thus different d-band centers (Figure \ref{architecture}c). To study the effects of the local environment on the properties of the active site, we curate, to the best of our knowledge, the largest dataset of multicomponent perovskite oxides by combining in-house DFT calculations with results compiled from the literature\cite{Castelli2012, Jacobs2018, Emery2016, Jain2013, Tran2022}. This dataset includes over 20,000 substituted bulk perovksite oxides up to quaternary, and over 10,000 surfaces cut along facets up to and including (555). The dataset additionally includes over 10,000 O$^*$, OH$^*$ and OOH$^*$ adsorption energies. For all bulk and surface structures we calculate per-site properties relevant to OER, including the oxygen 2p-band center and metal d-band center (Figure \ref{architecture}d), and magnetic moment (Figure \ref{architecture}g). For surface structures we additionally calculate  Bader charge (Figure \ref{architecture}f). Atomic vibration frequencies (Figure \ref{architecture}e), not calculated for our dataset but available from previous works\cite{PhononDatabase}, are also considered (Figure \ref{architecture}b). Using this dataset, we develop and train per-site graph-convolutional neural network machine learning models that are able to learn per-site properties (see Figure \ref{architecture}h) and ultimately OER intermediate binding energies (see Figure \ref{architecture}i) directly and automatically from atomic structure without the need for hand-crafted features.

\subsection{Learning compositional effects on per-site properties in bulk perovskites}

\begin{figure}
    \begin{center}
    \includegraphics[width=\textwidth]{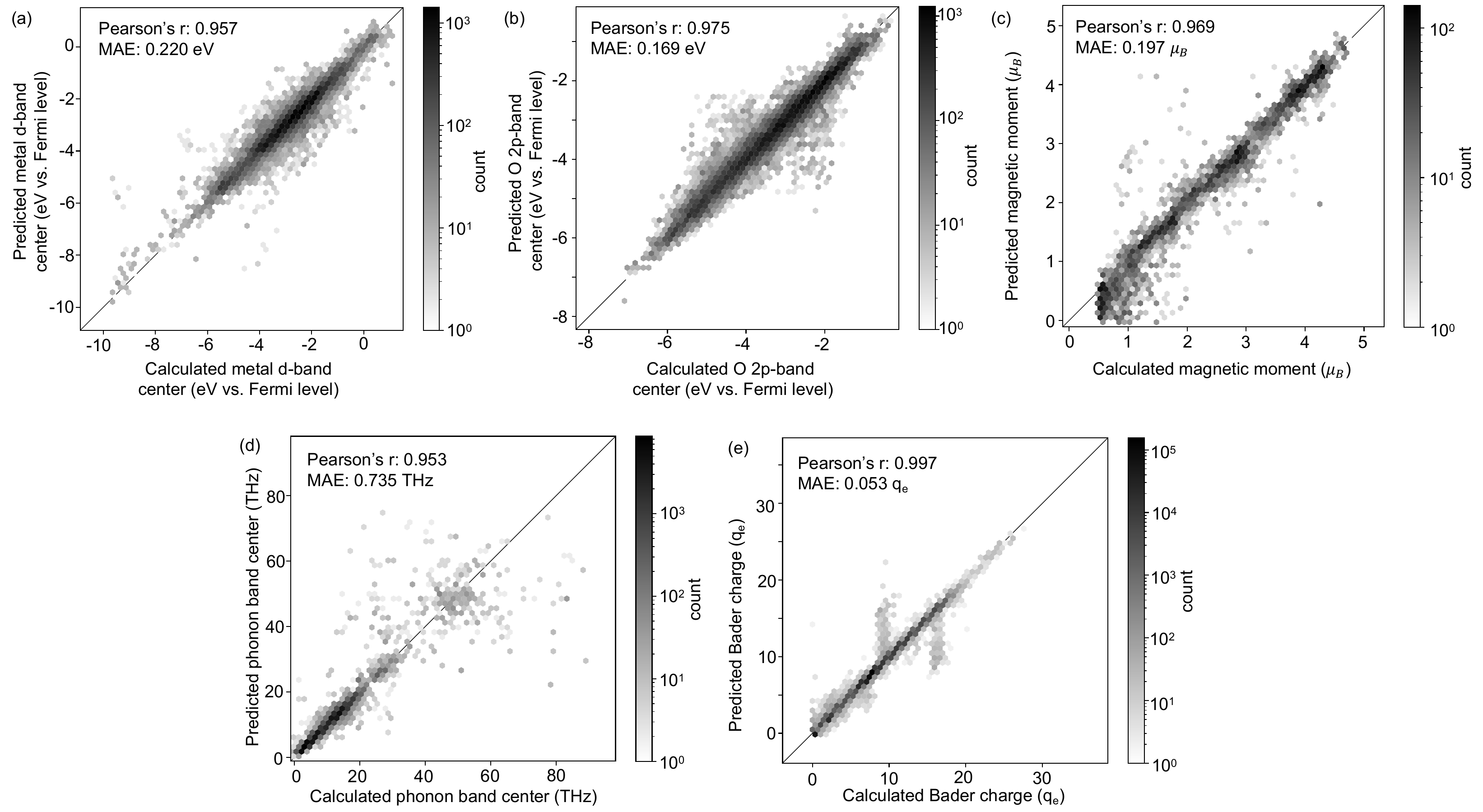}
    \end{center}
     \caption{per-site CGCNN test results on bulk atoms. Predicted vs. calculated (a) metal d-band center, (b) oxygen 2p-band center, (c) magnetic moment, (d) phonon band center and (e) Bader charge for all atoms in the test dataset.}
     \label{parity_bulk_SI}
\end{figure}

\begin{figure}
    \begin{center}
    \includegraphics[width=\textwidth]{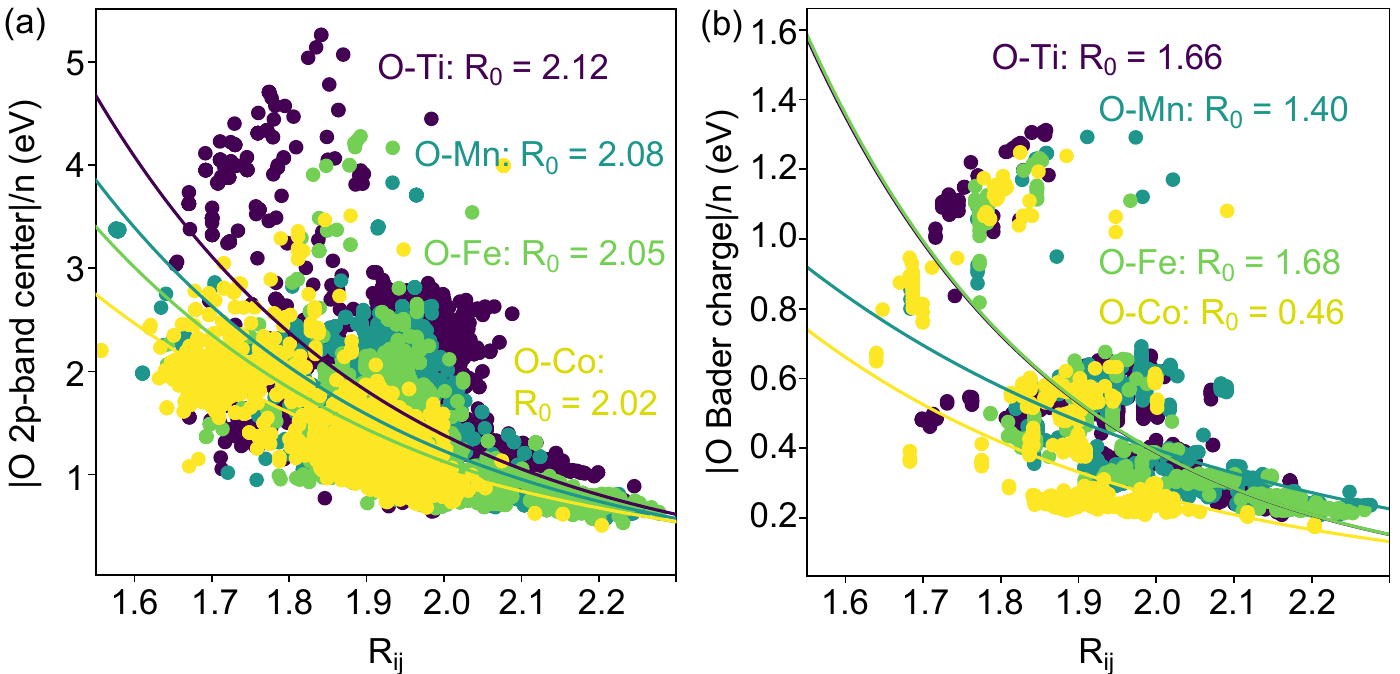}
    \end{center}
     \caption{Example fitting of bond-valence parameters for O-Ti, O-Mn, O-Fe, and O-Co for (a) O 2p-band center and (b) O Bader charge, where R$_{ij}$ is the distance between O and it's neighbor atom j and n is the coordination number of the oxygen.}
     \label{bond_valence}
\end{figure}

We show that by automatically encoding the local environment around each site, graph-based per-site models are able to capture the effect of chemical environment on per-site properties with high accuracy and without requiring hand-tuned features. We train and hyper-parameter tune several different site-aware deep learning models (see more details in Supplementary Information) on the bulk perovskites in our dataset to predict a number of DFT-derived atomic properties (metal d-band center, O 2p-band center and magnetic moment), which are functions of the geometry and composition of the local chemical environment of each atom. We additionally train these models on external datasets for predicting Bader charge\cite{Jain2013} and phonon band center\cite{PhononDatabase}, which are described in more detail in the Methods section. Performance is reported as Pearson's correlation coefficient on a held-out test-set for these per-site properties in Table \ref{per_site_table_bulk}, and parity plots for each property are reported in Figure SI\ref{parity_bulk_SI}. The per-site model based on Crystal Graph Convolutional Neural Network (CGCNN)\cite{Xie2018} and Polarizable Atom Interaction Neural Network (PAINN)\cite{Schutt2021} architectures have very similar performance. The best results are shown in bold in Table \ref{per_site_table_bulk}, and are 0.952, 0.969, 0.950, 0.962 and 0.997 for metal d-band center, oxygen 2p-band center, magnetic moment, phonon band center, and Bader charge, respectively. While for magnetic moment, Bader charge and phonon band center, element type alone maps well to the property with correlation coefficients of 0.904, 0.957, and 0.924, respectively, the site-projected metal d-band and oxygen 2p-band centers are highly dependent on the environment around the site as these properties are a result of orbital overlaps with neighboring atoms. Moreover, we consider predictions from the bond-valence method\cite{OKeefe1991}, where per-site properties are estimated from neighbor distances (P$_i$ = $\sum_{j}exp(\frac{R_0-R_{ij}}{B})$, having $R_{ij}$ as the bond distance between site i and neighbor j, $R_0$ as the fit for a given cation-anion pair, and B as an empirical constant typically set to 0.37\AA\cite{OKeefe1991}). Bond-valence method parameters were fit for each property using our dataset, and an example of this fitting for predicting oxygen 2p-band center and oxygen Bader charge is shown for O-Ti, O-Mn, O-Fe and O-Co bonds in Figure SI\ref{bond_valence}. Fitted bond-valence parameters are reported in Table SI\ref{bond_valence_parameters}. It is shown that R$_0$ for predicting oxygen Bader charge decreases from 2.12 to 2.08, to 2.05, to 2.02 for O-Ti, O-Mn, O-Fe and O-Co bonds, respectively, suggesting that oxygen 2p-band center is increasingly pushed down by oxygen-neighboring elements from left to right across the d-band. The bond valence method often outperforms simple element averages by accounting for neighboring atom types and distances. By encoding for several shells of neighbors through multiple graph convolutions, the per-site model is able to significantly outperform both element averages and the bond-valence method without the need for hand-tuned parameters (Table \ref{per_site_table_bulk}).

\begin{table}
  \caption{Model and benchmark test performance in terms of Pearson's correlation coefficients. Metal d-band center, oxygen 2p-band center, magnetic moment, and Bader charge are learned simultaneously while Bader charge and phonon band center are from separate datasets and learned individually. For magnetic moment, only magnetic atoms (with moments greater than 0.5 $\mu_B$) are considered and included in the test set statistics. Errors are calculated from three experiments with random weight initializations and test/train splits. The best performing model is highlighted in bold for each property. A correlation coefficient for O 2p-band center element average was not computed as it is not defined for comparison against a single value. The bond-valence method for magnetic moment is not reported, as the parameters were unable to be fit without high error.} 
  \label{per_site_table_bulk}
  \centering
  \begin{adjustbox}{width=\textwidth}
  \begin{tabular}{lccccc}
    \toprule
    \shortstack{per-site property} &  \shortstack{per-site \\ CGCNN} & \shortstack{per-site \\ PAINN} & \shortstack{bond-valence \\ method} & \shortstack{element average \\ in dataset}\\
    \midrule
    Metal d-band center  & \textbf{0.952 $\pm$ 0.003} & 0.939 $\pm$ 0.003  & 0.722 $\pm$ 0.002 & 0.515 $\pm$ 0.013 \\
    O 2p-band center  & \textbf{0.969 $\pm$ 0.010}   & 0.962 $\pm$ 0.003  & 0.588 $\pm$ 0.003 &  \texttt{--} \\
    magnetic moment & \textbf{0.950 $\pm$ 0.007}  & \textbf{0.947 $\pm$ 0.008}  & \texttt{--}  & 0.874 $\pm$ 0.030  \\
    \midrule
    phonon band center & 0.940 $\pm$ 0.011  & \textbf{0.962 $\pm$ 0.012} & 0.934 $\pm$ 0.012  & 0.918 $\pm$ 0.009 \\
    Bader charge & \textbf{0.997 $\pm$ 0.001} & 0.994 $\pm$ 0.001 & 0.821 $\pm$ 0.019 & 0.946 $\pm$ 0.013 \\
    \bottomrule
  \end{tabular}
  \end{adjustbox}
\end{table}

\begin{figure}
    \begin{center}
    \includegraphics[width=\textwidth]{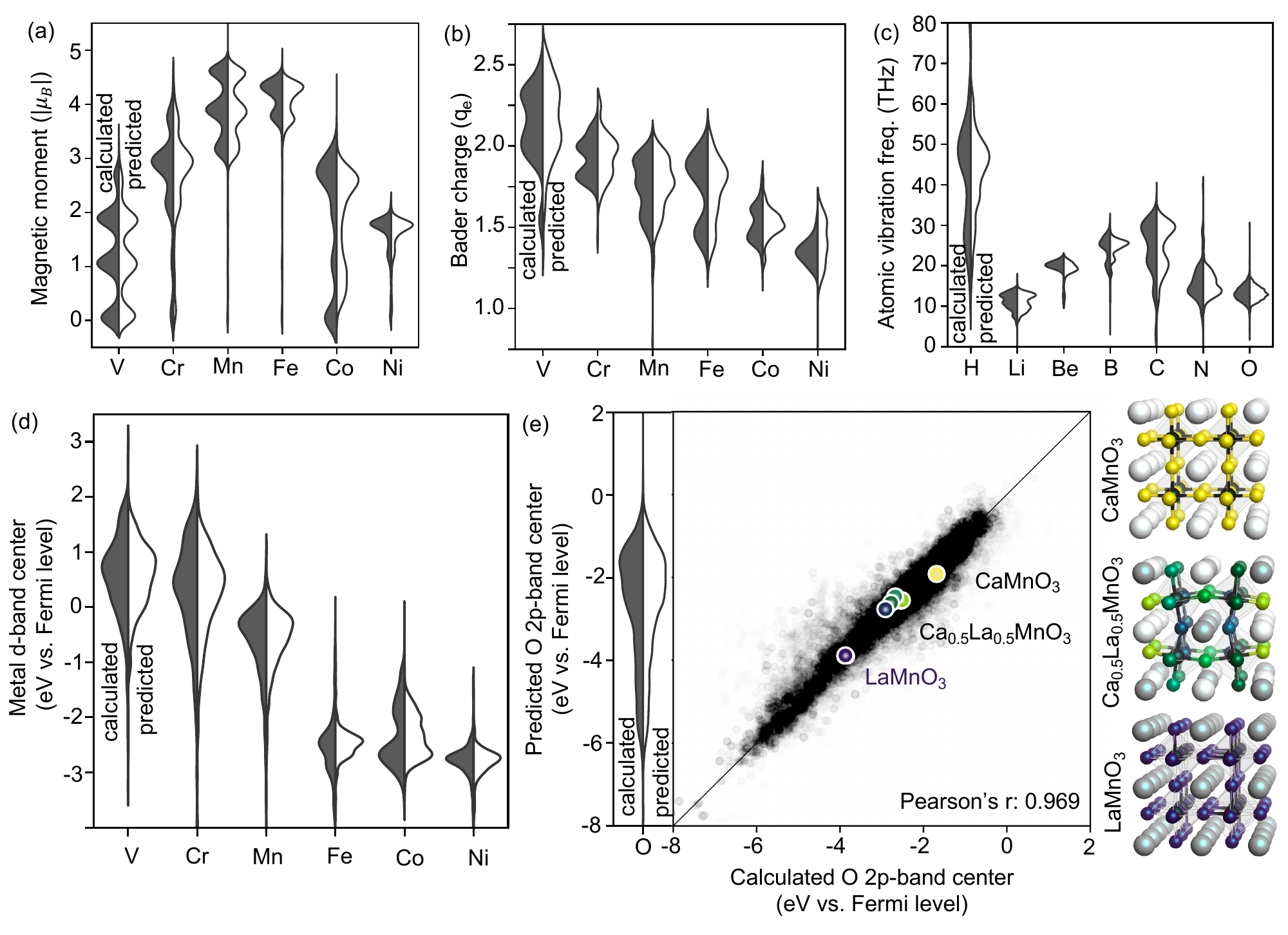}
    \end{center}
     \caption{Compositional effects of per-site properties. Predicted versus calculated per-site (a) magnetic moment, (b) Bader charge, (c) atomic vibration frequency and (d) metal d-band center as a function of element type going across the 3d band of the periodic table. (e) Predicted versus calculated oxygen 2p-band center showing several example compositions.}
     \label{model_results}
\end{figure}

We compare the calculated and predicted per-site materials properties across a held-out test set on an element-wise basis to highlight the ability of the per-site model to explicitly learn physical principles dictated by local environments. Calculated and predicted magnetic moments (Figure \ref{model_results}a), Bader charges (Figure \ref{model_results}b), atomic vibration frequencies (Figure \ref{model_results}c), metal d-band centers (Figure \ref{model_results}d), and oxygen 2p-band centers (Figure \ref{model_results}e) on a held-out test set are shown as a function of element for a sample of elements (V through Ni for metal d-band center, magnetic moment, and Bader charge; H through O for atomic vibration frequency; and O for oxygen 2p-band center). The models are shown to accurately capture both overall periodic trends (predictable from basic physical principles) as well as element-wise distributions (non-trivial effects of local environment). For example, the magnetic moments follow Hund's rules\cite{Kutzelnigg1996}: the number of unpaired electrons increases from V to Mn as separate d states are filled, then decreases from Mn to Ni as electrons co-fill d-states and spin is cancelled (Figure \ref{model_results}a). The model recreates that elements become more negatively charged going from V to Ni as electronegativity increases going from left to right along the periodic table, as reflected by decreasing Bader charge (Figure \ref{model_results}b). Per-site atomic vibration frequency is shown to generally decay for the first 10 elements in the periodic table as mass of the element increases (Figure \ref{model_results}c). Periodic trends in d-band center are also captured: as the number of electrons in the d-band increases across the periodic table, the d-band width must widen to maintain the Fermi level thus pushing the d-band center more negative\cite{Norskov2014} (Figure \ref{model_results}d). 

The per-site model accurately captures per-element distributions in material properties, which is remarkable as these distributions represent complex effects of substitution and local environment not obvious from basic physical principles alone. Additionally, the nodes in the calculated and predicted magnetic moment (Figure \ref{model_results}a) and Bader charge (Figure \ref{model_results}b) for each element represent different oxidation and spin states which are typically difficult to predict. The element-wise distributions in atomic vibration frequency (Figure \ref{model_results}c), metal d-band center (Figure \ref{model_results}d), and oxygen 2p-band center (Figure \ref{model_results}e) are even less obvious, as these spreads do not include nodes corresponding to discrete oxidation states. For example, the model captures that oxygen atoms in Ca$_{0.5}$La$_{0.5}$MnO$_3$ have a spread of 2p-band centers between -2.85 and -2.50 eV vs. Fermi (Figure \ref{model_results}e). Indeed, this range is an interpolation of the oxygen 2p-band centers of -1.67 and -3.87 eV vs. Fermi in CaMnO$_3$ and LaMnO$_3$, respectively. Notably, the 2p-band center is higher for oxygen atoms with more Ca neighbors and lower for oxygen atoms with more La neighbors. 

Despite the non-obvious nature of the site-properties discussed here, the presented models capture different local environments, match the calculated data with high accuracy, and are a powerful tool for tuning site properties through substitution effects.

\subsection{Learning facet and composition dependence of surface site properties}

\begin{figure}
    \begin{center}
    \includegraphics[width=\textwidth]{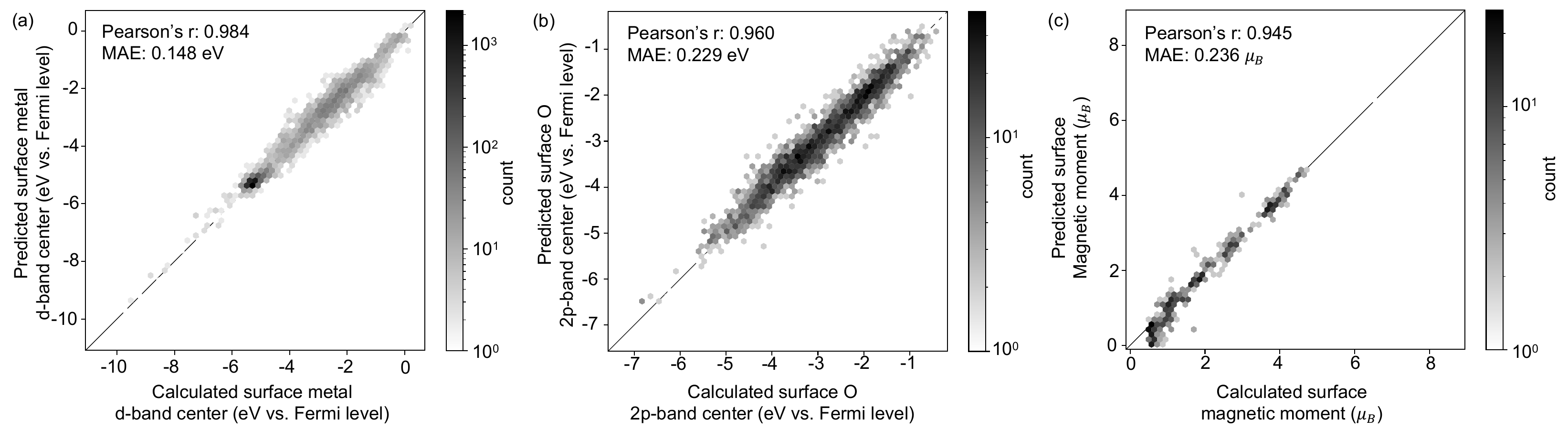}
    \end{center}
     \caption{per-site CGCNN test results on surface atoms. Predicted vs. calculated (a) metal d-band center, (b) oxygen 2p-band center, and (c) magnetic moment for all surface atoms in the test dataset.}
     \label{parity_surface_SI}
\end{figure}

Moving beyond bulk towards a more realistic representation of catalytic sites, we use the surfaces in our dataset to show the strong influence of surface  coordination and composition on per-site properties of surface atoms and active sites. The results for learning metal d-band center, oxygen 2p-band center, magnetic moment, and Bader charge of specifically surface atoms are summarized in terms of Pearson's correlation coefficient in Table \ref{per_site_table_surfaces}, and parity plots for each property are also reported in Figure SI\ref{parity_surface_SI}. The per-site graph neural network models are predictive of these properties with a correlation coefficient of 0.980, 0.955, 0.969, and 0.995, respectively, and are thus able to capture surface effects. The per-site CGCNN and PAINN models have similarly high performance, analogously to that presented for bulk atoms in Table \ref{per_site_table_bulk}. It is shown that for surface atoms the bond valence method loses some of its predictability, likely due to lack of accounting for undercoordination. We additionally compare the per-site model results to a baseline of the element average in the corresponding bulk structure, which is able to capture some composition effects but cannot account for local environment effects such as facet dependence (Table \ref{per_site_table_surfaces}). That the per-site model significantly outperforms this reasonable baseline shows that the model has successfully encoded local environment of surface atoms specifically. 

\begin{table}
  \caption{Model and benchmark test performance on surface atoms only in terms of Pearson's correlation coefficient. Errors are calculated by running 3 experiments with random weight initializations and test/train splits. The best performing model is highlighted in bold for each property. Element average in corresponding bulk was not available for Bader charges. The bond-valence method for predicting magnetic moment failed due to lack of sufficient fitting.} 
  \label{per_site_table_surfaces}
  \centering
  \begin{adjustbox}{width=\textwidth}
  \begin{tabular}{lccccc}
    \toprule
    \shortstack{surface \\ per-site property} &  \shortstack{per-site \\ CGCNN} & \shortstack{per-site \\ PAINN} & \shortstack{bond-valence \\ method} & \shortstack{element average in \\ corresponding bulk}\\
    \midrule
    Metal d-band center  & \textbf{0.980 $\pm$ 0.001} & 0.975 $\pm$ 0.002 & 0.512 $\pm$ 0.011 & 0.670 $\pm$ 0.025 \\
    O 2p-band center  & \textbf{0.955 $\pm$ 0.001}  & 0.924 $\pm$ 0.007 & 0.493 $\pm$ 0.021 &  0.582 $\pm$ 0.037 \\
    magnetic moment & 0.954 $\pm$ 0.005 & \textbf{0.969 $\pm$  0.008}   & \texttt{--} & 0.894 $\pm$ 0.008 \\
    Bader charge & \textbf{0.995 $\pm$ 0.000} & 0.992 $\pm$ 0.001 & 0.938 $\pm$ 0.003  & \texttt{--} \\
    \bottomrule
  \end{tabular}
  \end{adjustbox}
\end{table}

\begin{figure}
    \begin{center}
    \includegraphics[width=\textwidth]{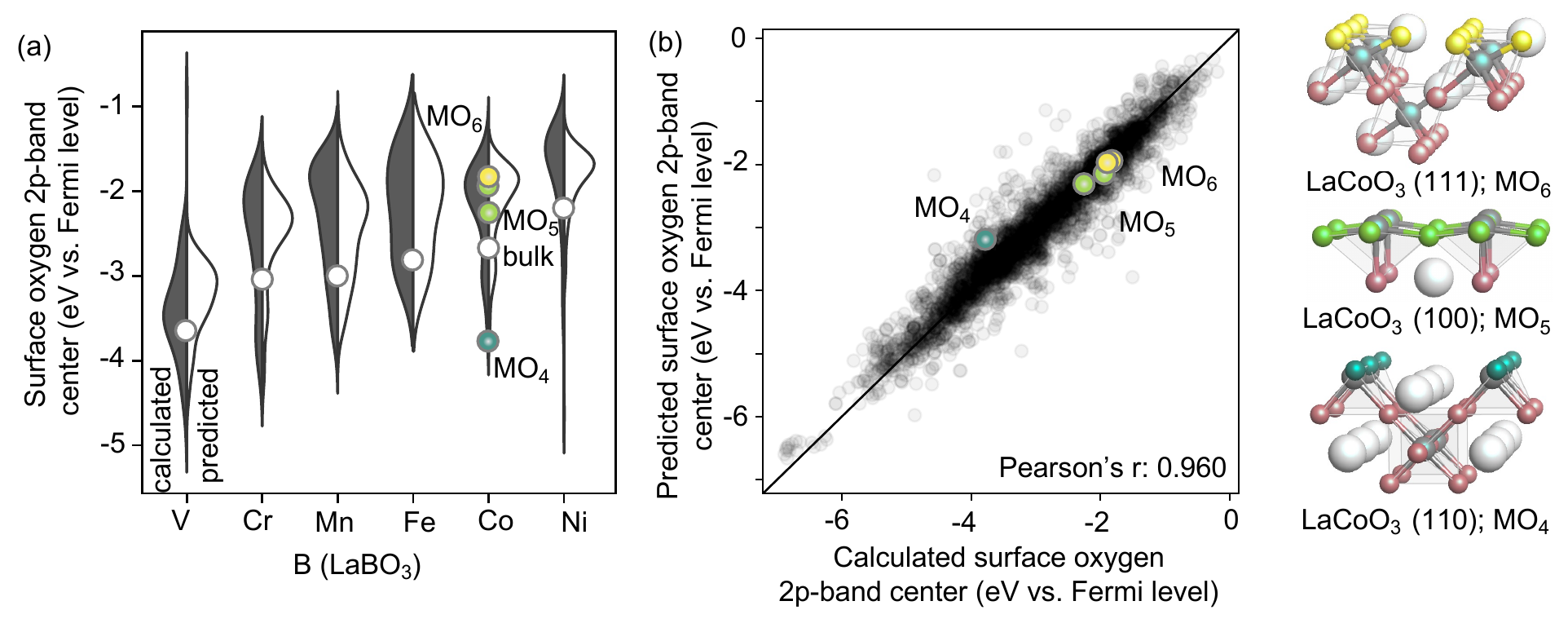}
    \end{center}
     \caption{Facet effects of per-site properties. (a) Calculated and predicted per-site 2p-band center of surface oxygens on LaBO$_3$ surfaces, for B going across the 3d band and cut along miller indexes up to (555). (b) Calculated versus predicted 2p-band center of surface oxygens for a held-out test set. The surface oxygens on the (100), (110) and (111) facets of LaMnO$_3$ are highlighted and the corresponding structures are shown.}
     \label{surface_descriptors}
\end{figure}

\begin{figure}
    \begin{center}
    \includegraphics[width=0.4\textwidth]{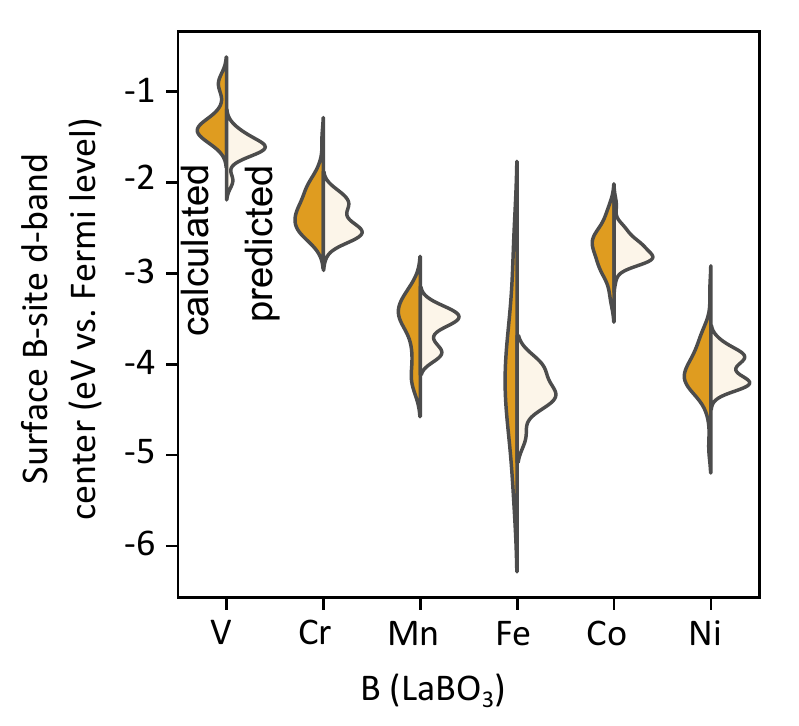}
    \end{center}
     \caption{Calculated vs. predicted d-band center of surface B-sites for a set of benchmark surfaces in the test set. LaBO$_3$ for B=V, Cr, Mn, Fe, Co and Ni with facets up to \(555\) are considered.}
     \label{d_band_surface}
\end{figure}

We compare the calculated and predicted per-site surface properties on an element-wise basis across a held-out test set of benchmark surfaces to further highlight the ability of the per-site model to explicitly learn composition and facet dependence. The predicted and calculated 2p-band centers of surface oxygens in LaBO$_3$ is compared as a function of B going across the 3d band of the periodic table from V to Ni, where the model has not been trained on any of these LaBO$_3$ materials (Figure \ref{surface_descriptors}a). Spreads in each of the violin plots are due solely to different surface coordinations in the dataset, which range from MO$_3$ to MO$_6$. Interestingly, the oxygen 2p-band center of surface oxygens varies significantly with facet, even more than the variation due to changing the B site. This wide spread is due to the significantly different coordination environments of facets that are close-packed versus not. Generally, the oxygen 2p-band centers of adsorbed oxygens become less negative as the active site is more highly coordinated (i.e. oxygens within MO$_6$ have higher 2p-band centers than those in MO$_4$). These spreads, which cannot be captured from bulk band center calculations (shown as a single point in white for each composition), provide compelling evidence for the need to account for surface coordination. In contrast to surface oxygen 2p-band centers, surface metal d-band centers are shown to have a much more limited dependence on facet and surface atom coordination, and are more strongly a function of composition  (Figure SI\ref{d_band_surface}). The model is able to capture both compositional effects (2p-band centers of surface-oxygen generally become less negative as B goes from left to right across the periodic table) as well as effects of faceting (calculated versus predicted distributions visibly match).

We highlight a few example surfaces in the test set, namely the (110), (100), and (111) facets of LaMnO$_3$ (or MO$_4$, MO$_5$, and MO$_6$ environments, respectively), to show that the 2p-band centers of adsorbed oxygens are most negative for the high coordination environment of (111) oxygen atoms (MO$_6$), and least negative for the lower coordination environment of (110) oxygen atoms (MO$_4$), and that the per-site model is able to capture this effect (Figure \ref{surface_descriptors}b). This trend also matches the findings of Rao et al.\cite{Rao2020}, who showed that bridge oxygen atoms on the edge-sharing (110) facet of unary RuO$_2$ had a less negative 2p-band center than the bridge oxygen atoms on the (101) corner-sharing facet.  We show that the per-site model is able to capture these facet effects, both by matching the distributions in Figure \ref{surface_descriptors}a and by correctly ordering the highlighted surface oxygen atoms in Figure \ref{surface_descriptors}b.

\subsection{Atom-by-atom design strategies for OER}

\begin{figure}
    \begin{center}
    \includegraphics[width=\textwidth]{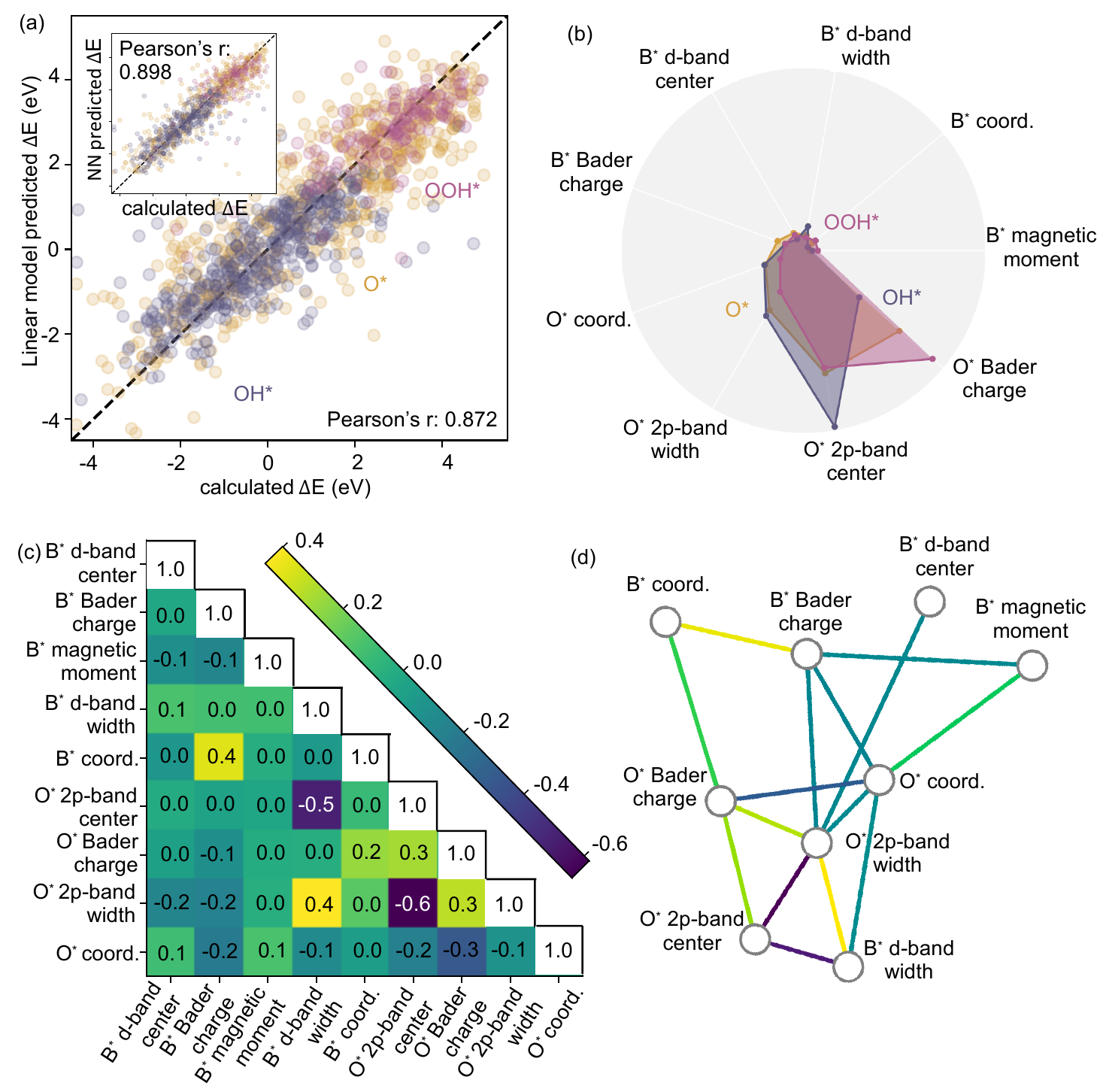}
    \end{center}
     \caption{Atom-by-atom design of OER energetics. (a) Binding energy of O$^*$, OH$^*$, and OOH$^*$ as a linear function of surface per-site properties. Binding energies compared to predictions from a neural network based on per-site properties are shown in the inset. (b) Feature importance of each per-site property with respect to binding energies. (c) Correlation matrix of surface per-site properties. (d) Surface per-site property correlation network with Bonferroni corrected p-values $>$ 0.01 filtered out. }
     \label{binding_energy}
\end{figure}

\begin{figure}
    \begin{center}
    \includegraphics[width=\textwidth]{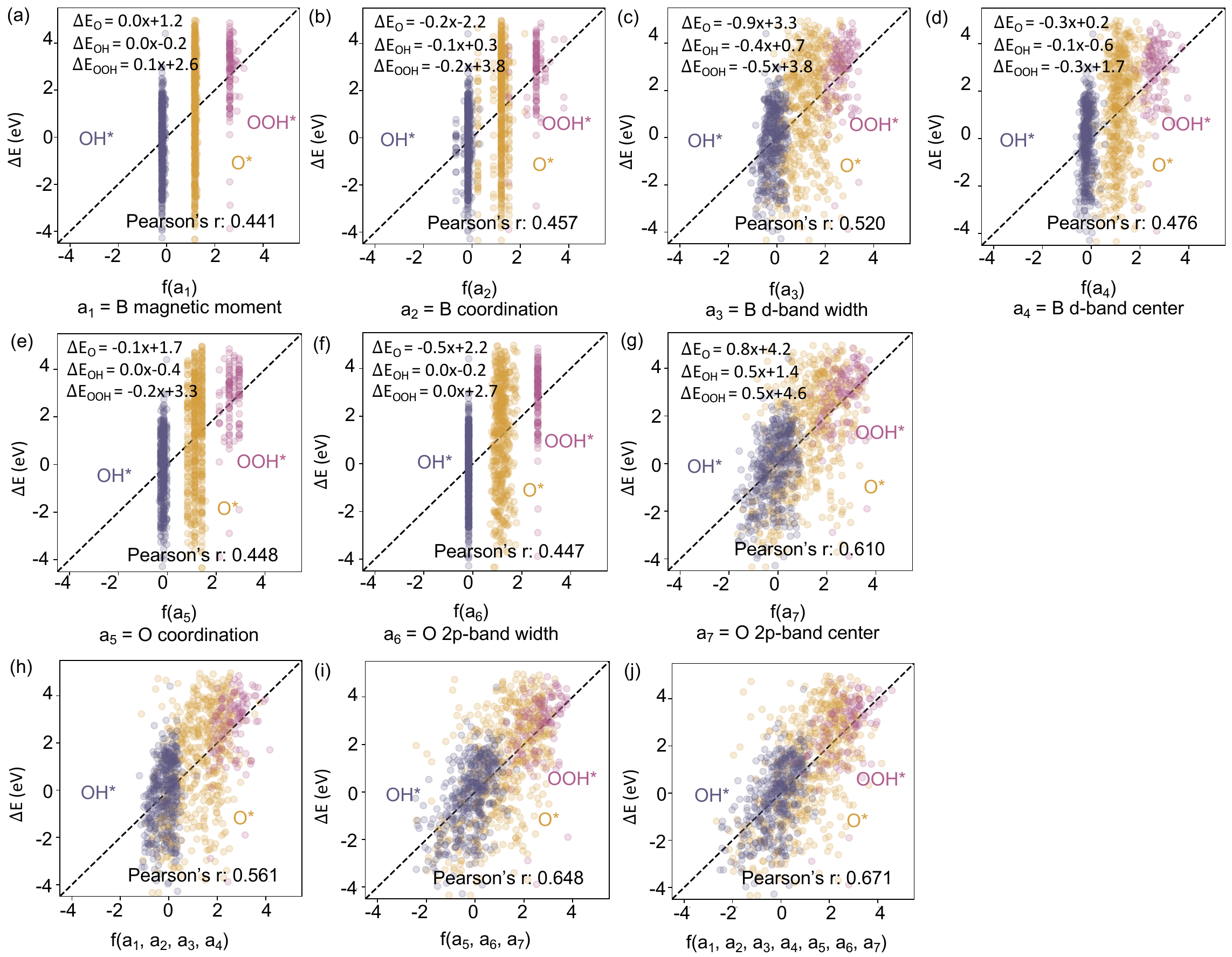}
    \end{center}
     \caption{Binding energies of O$^*$, OH$^*$ and OOH$^*$ as a linear function of bulk descriptors.}
     \label{bulk_descriptors}
\end{figure}

\begin{figure}
    \begin{center}
    \includegraphics[width=\textwidth]{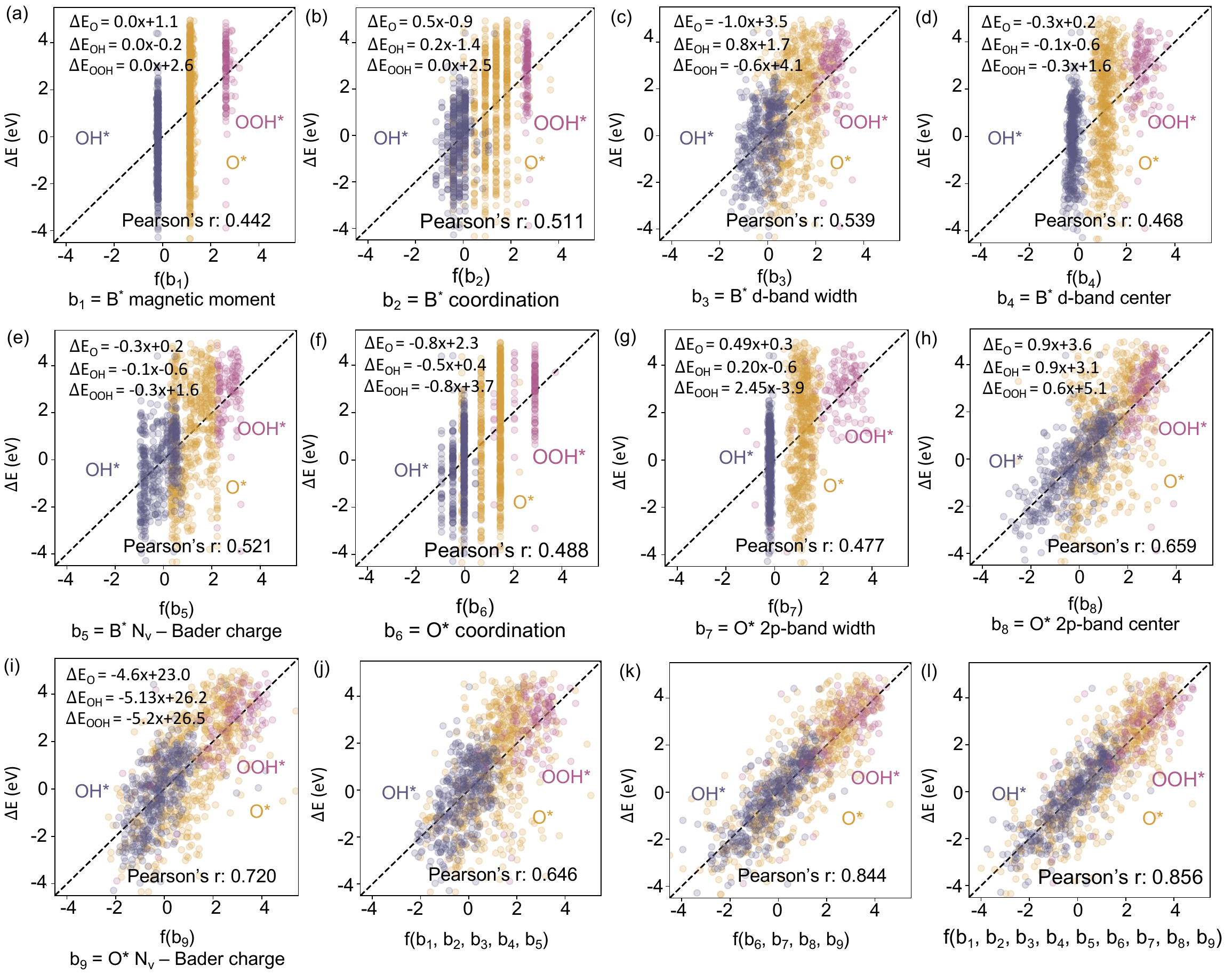}
    \end{center}
     \caption{Binding energies of O$^*$, OH$^*$ and OOH$^*$ as a linear function of per-site descriptors of adsorbed oxygen (O$^*$) and/or the active site (B$^*$).}
     \label{per-site_surface_descriptors}
\end{figure}

We next demonstrate the ability of per-site descriptors to predict DFT-calculated binding energies of OER intermediates, and ultimately catalytic activity. We fit a multi-linear function of per-site properties of the active site (magnetic moment, Bader charge, d-band center, d-band width, and coordination number) and of the adsorbed oxygen (Bader charge, 2p-band center, 2p-band width, and coordination number) to OER intermediate binding energies. The predicted versus calculated binding energies are shown in Figure \ref{binding_energy}a, with the linear model achieving excellent predictive power with a correlation coefficient of 0.872. These simple linear relationships between binding energy and per-site properties capture most of the variation in the data. A complex, non-linear model combining a per-site graph neural network and per-site properties (Figure \ref{binding_energy}a inset for results on a held-out test set) moderately outperforms the linear model with a correlation coefficient of 0.898. These results are in agreement with and extend beyond previous studies\cite{Dickens2019, Hong2015, Hwang2017, Jacobs2019}, which have shown that OER intermediate binding energies correlate to bulk oxygen 2p-band center\cite{Hong2015,Jacobs2019} or to adsorbed oxygen 2p-band center\cite{Dickens2019,Hwang2017}. We show these same correlations between binding energies and bulk descriptors\cite{Hong2015, Jacobs2019}, and between binding energies and individual per-site descriptors\cite{Hwang2017,Dickens2019} in Figure SI\ref{bulk_descriptors} and Figure SI\ref{per-site_surface_descriptors}. Combining these models with the above-presented per-site property prediction models, and with models that predict relaxed structure\cite{Chen2022,Deng2023}, would enable end-to-end learning of binding energies without the need for DFT calculations. Moreover, unlike these previous efforts\cite{Dickens2019, Hong2015, Hwang2017, Jacobs2019} which have only included the $(001)$ facet of perovskites, our work includes facets up to $(555)$ and multimetal complex perovskites up to quaternary. 

By leveraging multiple per-site properties of the active site and adsorbed oxygen simultaneously, we are able to improve upon these previous correlations (e.g. correlation coefficient of 0.61 for bulk oxygen 2p-band center\cite{Hong2015,Jacobs2019}, 0.66 for surface oxygen 2p-band center\cite{Dickens2019,Hwang2017}, and 0.87 for the linear model) as several of the per-site properties are found to simultaneously and independently influence the binding energy of OER intermediates. A polar plot of the relative importance of each per-site property assigned by normalizing the coefficients of the linear model is shown in Figure \ref{binding_energy}b. The two most important per-site properties are the 2p-band center of the adsorbed oxygen (in agreement with previous work\cite{Dickens2019, Lee2011, Hong2015}), and Bader charge of adsorbed oxygen, which is similar to e$_g$ filling\cite{Suntivich2011} and has been shown to be useful for both prediction of OER activity and metal oxide formation energy\cite{Suntivich2011,Craig2023,Hong2016}. The relative importance of these two per-site properties differs among adsorbates. For example, for $OH^*$, the least oxidizing adsorbate, the 2p-band center is the most influencial while for $OOH^*$, the most oxidizing adsorbate, Bader charge is the most influential. For $O^*$, these two descriptors and the oxygen 2p-band width are of more equal weight, as shown in Figure \ref{binding_energy}b. For all three adsorbates, the remaining parameters studied in this work (i.e. coordination number of adsorbed oxygen and Bader charge, d-band center, d-band width, coordination number, and magnetic moment of the active site) are found to be of limited importance for predicting DFT-derived OER binding energetics.

\begin{figure}
    \begin{center}
    \includegraphics[width=\textwidth]{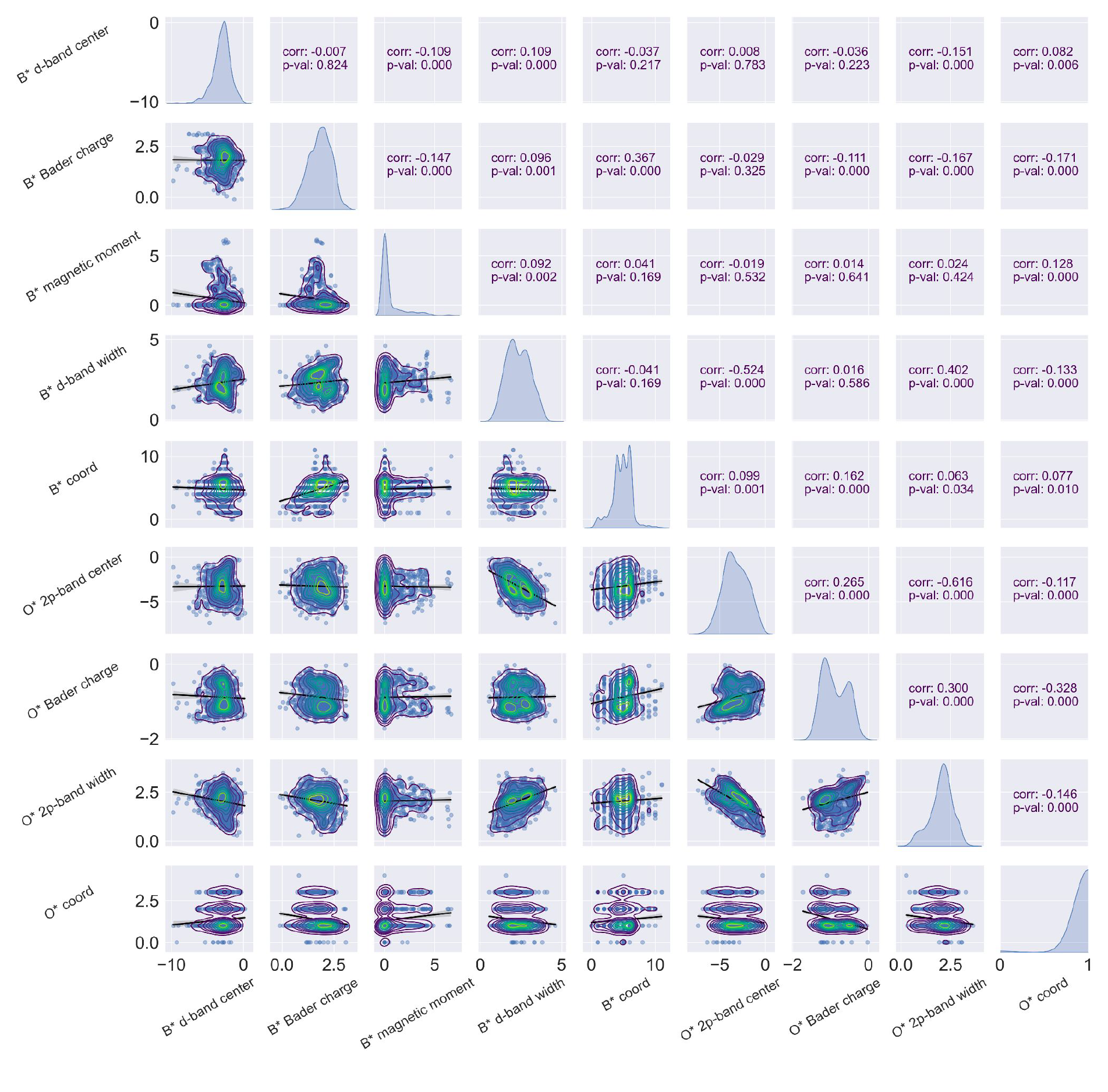}
    \end{center}
     \caption{Pairwise joint distributions overlaid with best fit lines (lower triangle) and marginal distributions (diagonal) of surface per-site descriptors. Pearson correlations and multi-hypothesis Bonferroni corrected p-values are also shown (upper triangle).}
     \label{descriptor_correlations_and_pvalues}
\end{figure}

A correlation matrix of the per-site properties is shown in Figure \ref{binding_energy}c. While the 2p-band width and 2p-band center of adsorbed oxygen are negatively correlated with a coefficient of -0.6, most of the other per-site properties are not found to correlate strongly with any of the others. Further support comes from the sparsity (median degree of three) of the correlation network after correcting for multiple hypothesis testing and filtering for connections with greater than 0.01 p-value (Figure \ref{binding_energy}d). Pairwise p-values were calculated across surface per-site descriptors and normalized by total comparison count as directed by the Bonferroni method for multiple hypothesis correction (Figure SI\ref{descriptor_correlations_and_pvalues}). Of interest, the metal d-band center does not significantly correlate with other metal descriptors suggesting it cannot be easily tuned by changing metal coordination. Further, the metal d-band center is of low feature importance to binding energy (Figure \ref{binding_energy}b) counter to common metal catalyst behavior \cite{Sun2020}. Most notably, the 2p-band center and Bader charge of adsorbed oxygen, the two per-site properties that most influence the binding energies, are found to be only weakly correlated with a correlation coefficient of 0.3 (Figure \ref{binding_energy}c), suggesting that these two per-site properties can be tuned independently.






\begin{figure}
    \begin{center}
    \includegraphics[width=\textwidth]{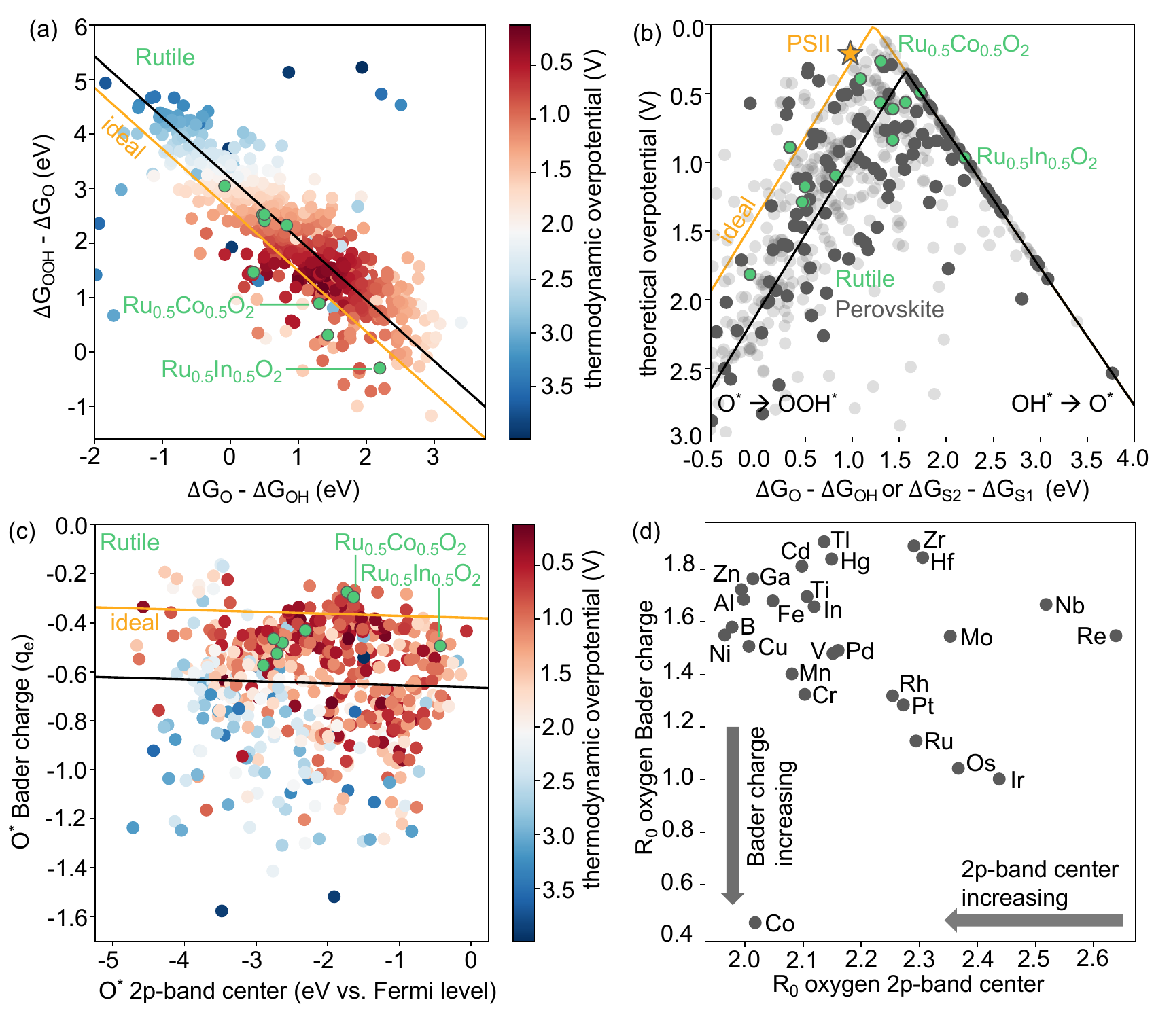}
    \end{center}
     \caption{Atom-by-atom design of activity. (a) Thermodynamic overpotential as a function of $\Delta G_O$ - $\Delta G_{OH}$ and $\Delta G_{OOH}$ - $\Delta G_{O}$. Well-known scaling relations (black)\cite{CalleVallejo2013} and ideal scaling (orange) are also plotted. Substituted rutile oxides of interest are highlighted in green. (b) Volcano plot of the theoretical overpotential as a function of $\Delta G_O$ - $\Delta G_{OH}$. For PSII, the energy difference between the S$_2$ and $S_1$ states is used in place of $\Delta G_O$ - $\Delta G_{OH}$. (c) Thermodynamic overpotential is shown as a function of adsorbed oxygen 2p-band center and Bader charge. (d) Fitted R$_0$ values for predicting oxygen Bader charge, vs. R$_0$ values for predicting oxygen band center as a function of neighbor atom type.}
     \label{contour_plots}
\end{figure}

\begin{figure}
    \begin{center}
    \includegraphics[width=\textwidth]{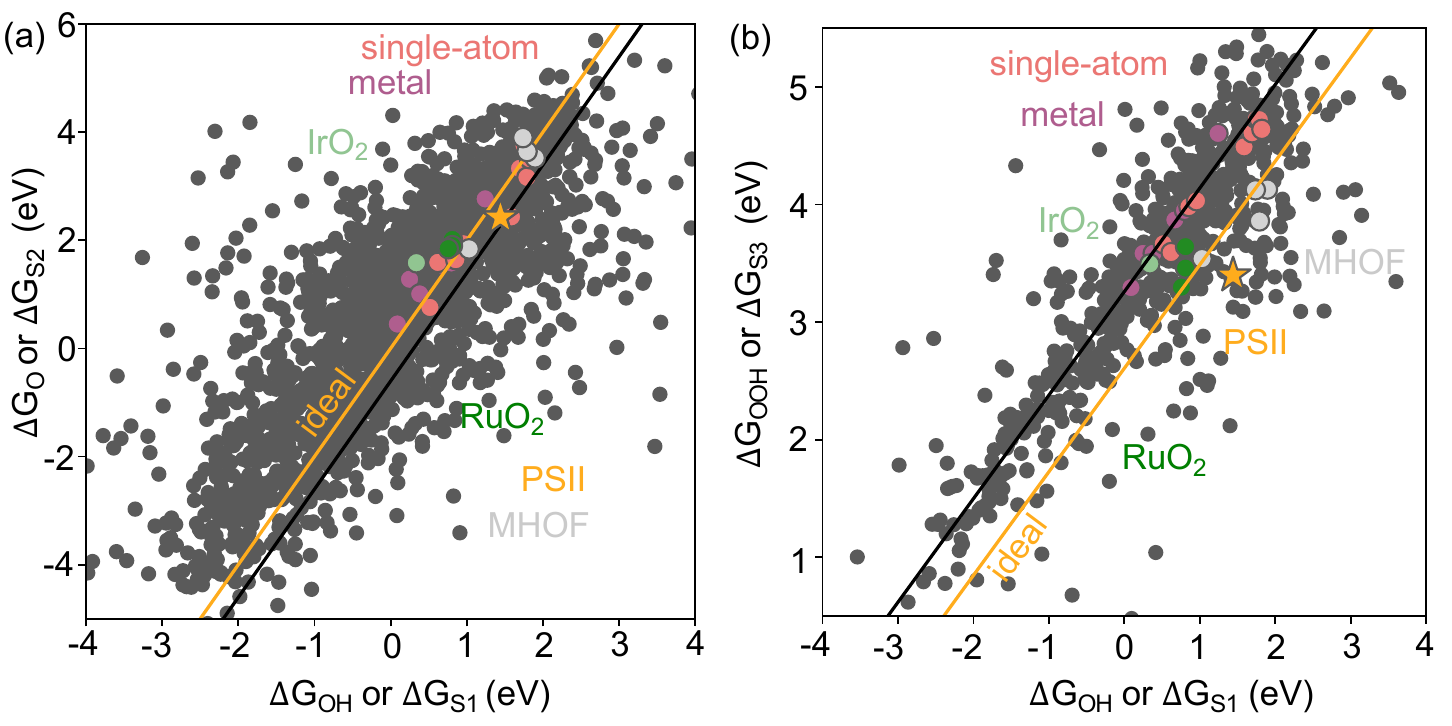}
    \end{center}
     \caption{(a) Scaling relations between $\Delta G_{OH^*}$ and $\Delta G_{O^*}$ for metals (purple circles)\cite{Kulkarni2018}, single atom catalysts (pink circles)\cite{Deng2019}, metal hydroxide-organic frameworks (MHOFs, light grey circles)\cite{Yuan2022}, photosystem II (orange star)\cite{Siegbahn2013}, RuO$_2$ facets\cite{Rao2020}, IrO$_2$ facets (unpublished data, light green circles)\cite{Gauthier2017}, and multicomponent oxides considered in this study (dark grey circles). For photosystem II, the free energy of S1, S2 and S3 is used in place of O$^*$, OH$^*$ and OOH$^*$ binding energies respectively. (b) Scaling relations between $\Delta G_{OH^*}$ and $\Delta G_{OOH^*}$. Scaling relations from Calle-Vallejo et al. are shown in black, and ideal scaling in orange\cite{CalleVallejo2013}.}
     \label{scaling_relationships}
\end{figure}

We compare the energetics for OER across the surfaces dataset to identify the most promising catalysts for OER, and develop design strategies for navigating this landscape as a function of per-site properties. The binding energies of OER intermediates (O$^*$, OH$^*$, and OOH$^*$) are known to be linearly dependent\cite{Man2011}, fundamentally limiting the OER overpotential to above 0.37 eV and preventing catalyst design via DFT, assuming proton-copuled electron transfer\cite{Rossmeisl2007}. For example, $\Delta E_{OH}$ is known to scale with $\Delta E_{O}$ with a slope of 0.5 eV and intercept of 0.44 eV, and $\Delta E_{OOH}$ scales with $\Delta E_{OH}$ with a slope of 0.88 eV and intercept of 3.21 eV\cite{CalleVallejo2013}. These well-known scaling relationships\cite{CalleVallejo2013,Man2011} compare well with our calculated values of $\Delta G_{O}-\Delta G_{OH}$ and $\Delta G_{OOH}-\Delta G_{O}$ for the dataset of complex multimetal perovskites studied in this work (Figure \ref{contour_plots}a), with data-points colored by their calculated theoretical overpotential. An additional scaling relationship, the "ideal" scaling, is plotted in orange in Figure \ref{contour_plots}a, where the well-known slope remains the same but the coefficient is adjusted such that scaling passes through 0 V overpotential. Interestingly, the spread in the perovksite oxides dataset is wide enough to touch the ideal scaling line due to the large diversity in composition and facet. We highlight several substituted rutile oxides of interest, which deviate from typical scaling and are closer to the ideal line due to substitution effects on the per-site properties and thus on OER binding energetics. Additional scaling relationships between $\Delta G_{OH}$ and $\Delta G_{O}$, and between $\Delta G_{OH}$ and $\Delta G_{OOH}$ are shown in Figure SI\ref{scaling_relationships}, where the materials in this work are additionally compared to previous efforts on metals (purple circles)\cite{Kulkarni2018}, single-atom catalysts (pink circles)\cite{Deng2019}, RuO$_2$ facets (dark green)\cite{Rao2020}, IrO$_2$ facets (light green)\cite{Gauthier2017}, metal hydroxide-organic frameworks (MHOFs, light grey)\cite{Yuan2022}, and PSII (orange star)\cite{Siegbahn2013}. While previously studied single-atom and metal catalysts fall more tightly on the typical scaling line (black, Figure SI\ref{scaling_relationships}), the oxides studied in this work show more deviation from the well-known scaling relations due to the large variation in both element substitution and faceting. RuO$_2$ facets from the work of Rao et al.\cite{Rao2020} are also shown to deviate from this scaling, due to corner-sharing versus edge sharing effects as well as stabilization of OOH$^*$ by neighboring coordinately unsaturated oxygen. The effect of this deviation from scaling is that some surfaces studied in this work are able to outperform the well-known OER thermodynamic overpotential limit, with overpotentials as low as 0.21 V La$_{0.5}$Sr$_{0.5}$CoO$_3$ (211). We note that the deviation from scaling for some of these perovskites is even as extreme as the deviation for photosystem II (PSII)\cite{Siegbahn2013}(Figure SI\ref{scaling_relationships}) The typical activity volcano for theoretical potential as a function of $\Delta G_{O}-\Delta G_{OH}$ is also shown (Figure \ref{contour_plots}b), including standard and ideal scaling (black and orange lines, respectively), perovskites (dark grey), and rutile oxides (green). Perovksite oxides in our dataset that are unstable (energy above hull of the corresponding bulk structure above 0.3 eV) are semi-transparent while stable perovskites (energy above hull below 0.3 eV) are fully opaque. Energetics for PSII\cite{Siegbahn2013} are additionally compared to the perovskites and rutiles in this work, and several materials with overpotentials approaching that of PSII are plotted in Figure \ref{contour_plots}b and listed in Table \ref{promising_materials}.

A new version of the activity volcano as a function of relevant per-site properties rather than binding energies is shown (Figure \ref{contour_plots}c), to show the physical origin of the deviation from the standard scaling relationships. The 2p-band center and Bader charge of adsorbed oxygen are considered, because these properties have the strongest influence on binding energies, are not correlated, and can be tuned independently to break the scaling relationships. While the 2p-band center of adsorbed oxygen is a good descriptor for traversing the standard scaling line, by simultaneously adjusting the Bader charge of the adsorbed oxygen closer to 0 q$_e$, it is possible to move toward ideal scaling. We compare the $R_0$ fitted bond-valence parameters for how transition metal neighbors affect oxygen Bader charge and oxygen 2p-band center, to build a map for navigating per-site phase space by adjusting perovskite composition (Figure \ref{contour_plots}d). It is shown that as the atomic number of neighboring atoms increases, the $R_0$ fitted parameter for predicting oxygen 2p-band center increases and the oxygen 2p-band center therefore decreases. In contrast, the $R_0$ fitted parameter for predicting oxygen Bader charge given the neighbor atoms decreases (and therefore oxygen Bader charge increases) going from left to right across the periodic table. This map helps guide researchers for what substitutions to make in a perovskite to modify the per-site properties of adsorbed oxygen, and therefore to tune the catalytic activity. For example, to have moderate oxygen 2p-band center and an oxygen Bader charge close to 0, substituting materials with cobalt, ruthenium, or iridium is a good choice. This matches with our intuition, as many promising perovskites for OER contain these metals\cite{Hwang2017,Hong2016}. We compare Ru$_{0.5}$In$_{0.5}$O$_3$ and Ru$_{0.5}$Co$_{0.5}$O$_3$ to illustrate this point. Replacing neighboring In with Co slightly decreases the R$_0$ for oxygen 2p-band center, while dramatically decreasing the R$_0$ for oxygen Bader charge (Figure \ref{contour_plots}d). By swapping out In (Ru$_{0.5}$In$_{0.5}$O$_3$) for Co (Ru$_{0.5}$Co$_{0.5}$O$_3$), the 2p-band center increases slightly (moving toward the peak along the scaling line in Figure \ref{contour_plots}c) and the Bader charge of adsorbed oxygen also increases. The result is that Ru$_{0.5}$Co$_{0.5}$O$_3$ has a much higher activity than Ru$_{0.5}$In$_{0.5}$O$_3$. This example illustrates a strategy for navigating activity space by tuning per-site properties through composition tuning. Additional strategies could involve facet engineering, which modulates the 2p-band center of adsorbed oxygen while having less effect on the Bader charge\cite{Rao2020}.

\begin{figure}
    \begin{center}
    \includegraphics[width=0.5\textwidth]{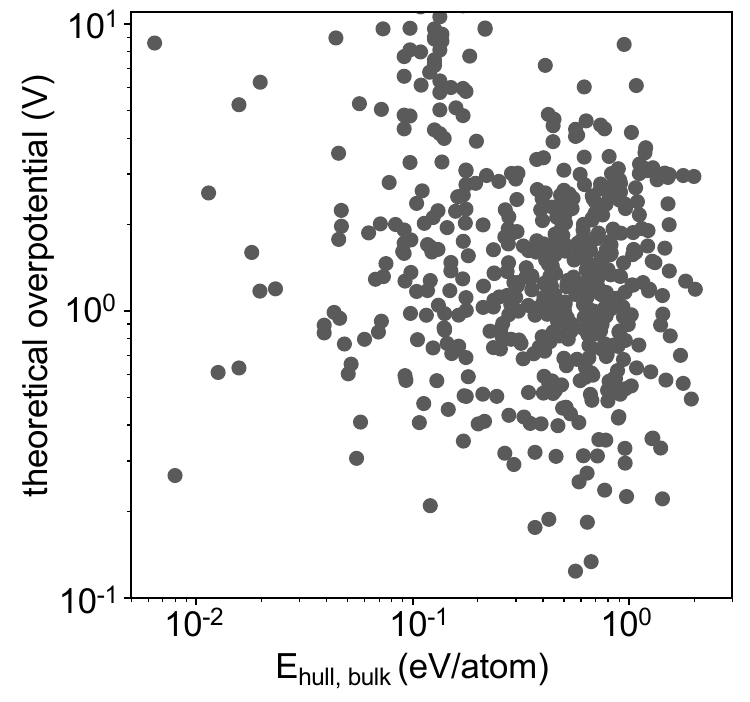}
    \end{center}
     \caption{Theoretical overpotential (V) vs. energy above hull of the corresponding bulk structure.}
     \label{ehull}
\end{figure}

Finally, the thermodynamic overpotential as well as the energy above hull of the corresponding bulk structure are used to screen the dataset for materials showing promising activity and stability (see Figure SI\ref{ehull}), and the most promising stable (energy above hull less than 0.3 eV) and active (thermodynamic overpotential below 0.6 V) materials are summarized in Table \ref{promising_materials}. While many of the interesting materials identified and listed in table \ref{promising_materials} have not yet, to the best of our knowledge, been tested for OER activity, several identified materials (such as La$_{0.5}$Sr$_{0.5}$CoO$_3$, LaNiO$_3$, PrNiO$_3$, and La$_{0.75}$Sr$_{0.25}$MnO$_3$) have already been tested and shown to be promising OER catalysts\cite{Hong2016, Suntivich2011}. Although the dataset curated and analyzed in this work includes over 10,000 binding energies, the phase space of multicomponent perovskite surfaces remains far from exhausted and we remain hopeful that many more promising perovskite materials may yet to be discovered. Toward this end, future work could include pairing per-site descriptors with effective strategies for intelligent search of the remaining phase space, such as Bayesian optimization\cite{Pedersen2019}. Also of interest are high entropy oxide systems with five or more metals, where the phase space is even more daunting and rich with opportunities for catalyst optimization\cite{Svane2022,Nguyen2021} through control of chemical ordering\cite{Peng2023}.

\begin{table}
  \caption{Materials in the dataset with bulk energy above hull below 0.3 eV/atom and thermodynamic overpotential below 0.6 V. Materials are sorted by thermodynamic overpotential, showing lowest to highest.}
  \label{promising_materials}
  \centering
  \begin{adjustbox}{width=0.5\textwidth}
  \begin{tabular}{lccc}
    \toprule
    chemical formula & miller index & \shortstack{thermodynamic \\ overpotential (V)}  \\
    \midrule
    La$_{0.5}$Sr$_{0.5}$CoO$_3$  & (211) & 0.21 \\ 
    Co$_{0.5}$Ru$_{0.5}$O$_2$ & (110) &  0.27 \\
    CaAuO$_3$ & (010) & 0.29 \\
    EuMn$_3$ & (010) & 0.31 \\
    LaBeO$_3$ & (111) & 0.32 \\
    KSbO$_3$ & (101) & 0.35 \\
    CaBiO$_3$ & (010) & 0.40 \\
    SrRhO$_3$ & (010) & 0.41 \\
    LiNbO$_3$ & (010) & 0.41 \\
    MnVO$_3$ & (010) & 0.41 \\
    LiRhO$_3$ & (110) & 0.43 \\
    CaVO$_3$ & (001) & 0.45 \\
    BaRhO$_3$ & (010) & 0.46 \\
    La$_{0.75}$Sr$_{0.25}$MnO$_3$ & (100) & 0.50 \\
    AgTeO$_3$ & (010) & 0.50 \\
    BaMnO$_3$ & (010) & 0.51 \\
    BaCoO$_3$ & (010) & 0.51 \\
    BiScO$_3$ & (010) & 0.57 \\
    LaNiO$_3$ & (211) & 0.57 \\
    PrNiO$_3$ & (320) & 0.59 \\

    \bottomrule
  \end{tabular}
  \end{adjustbox}
\end{table}

\section{Conclusions}

We develop a machine learning model based on graph-convolutional neural networks that is able to predict per-site properties of interest to catalysis from structure, including site-projected oxygen 2p-band center, metal d-band center, magnetic moment, phonon band center, and Bader charge. Using a dataset of over 19,000 multimetal complex bulk perovskites and 10,000 surface perovskites cut up to a miller index of (555), we show that the model is able to capture composition and substituting effects as well as surface faceting effects on per-site properties. We leverage per-site properties for predicting binding energies of OER intermediates, and demonstrate how Bader charge and 2p-band center of adsorbed oxygen work independently to tune the binding energies of OER intermediates outside of the typical scaling relations. A mapping between composition space and per-site property space is developed. Finally, we compare OER energetics on this largest yet known dataset of perovskite oxides, highlighting several "enzyme-like" active sites of interest. This work demonstrates the use of per-site properties, which enable atom-level design of the active site and, coupled with other strategies such as Bayesian optimization, may help close the gap between heterogeneous and enzymatic catalyst performance.

\section{Experimental procedures}

\subsection{Lead contact}

Further information and requests for resources should be directed to and will be fulfilled by the lead contacts, Rafael Gomez-Bombarelli, \texttt{rafagb@mit.edu} and Yang Shao-Horn, \texttt{shaohorn@mit.edu}.

\subsection{Materials availability}

This study did not generate any new materials.

\subsection{Data and code availability}

All code and datasets from this study will be made available at \url{https://github.com/learningmatter-mit/atom_by_atom}.

\subsection{Model details}

In this work, we implement machine learning models to both predict per-site properties directly from structure (see Figure \ref{architecture}h) and leverage these properties to predict OER intermediate binding energies (see Figure \ref{architecture}i).

Per-site graph-based neural network models (extensions of previous models\cite{Xie2018,Schutt2021}) were simultaneously trained on several calculated per-site properties including metal d-band center, oxygen 2p-band center, Bader charge, magnetic moment, and phonon band center. Convolutional layers are used to automatically featurize the structure based on local environments and a linear layer is used to convert this featurization to predicted per-site properties. Although all models in this study are trained on DFT-relaxed structures, we propose that an extension to unrelaxed structures will be possible using cheap surrogate models for predicting relaxed structure from unrelaxed structure\cite{Chen2022,Deng2023}. 

For all properties, we perform a benchmark comparison between per-site modified versions of CGCNN\cite{Xie2018} and PAINN\cite{Schutt2021}. Per-site CGCNN is initialized with physics-informed, curated atomic descriptors which improves generalizability and performance, especially with smaller datasets. For each per-site property in bulk structures, the per-site model performance is additionally compared against a baseline of element average across the dataset. For surface atoms, the per-site model performance is additionally compared against a baseline of the element average in the corresponding bulk structure, and to the bond-valence method. 

Linear models and neural networks were implemented to predict OER intermediate binding energies from calculated per-site properties. The linear model was regressed on calculated Bader charges, d-band centers, 2p-band centers, magnetic moments, and coordination numbers of surface active sites and adsorbed oxygens. A neural network architecture was also developed, using these properties appended to learned embeddings from the per-site graph-based models as input. This input was passed though graph convolutional layers and a final pooling to predict binding energies.

All results presented are on held-out test data. The models are trained on 60\% of the data, validated on 20\%, and tested on 20\%. Random train-validation-test splits were used in all cases. Model weights were chosen based on optimal validation set performance. Error ranges reflect statistics from three different experiments, where model weights and train-test splits are seeded randomly. Hyperparameter tuning was performed on all models using SigOpt\citep{sigopt-web-page}. All models were run on Nvidia GeForce RTX 2080 Ti GPUs.

\subsection{Datasets}

Multicomponent bulk perovskite oxides were compiled from the literature
\cite{Castelli2012, Jacobs2018, Emery2016, Jain2013} and augmented by in-house DFT calculations, resulting in over 19,000 bulk perovskite structures. DFT calculations were performed using the Vienna ab initio simulation package (VASP) and matching the default calculation parameters in Materials Project including pseudopotential and U\cite{Jain2013}. Unary, binary, ternary, and quaternary structures up to unit cell size 2x2x2 were considered. The 2p-band and d-band center values of each atom were obtained by projecting the density of states onto each site and integrating between -10 below and 2 above the Fermi level. Magnetic moments were obtained directly from VASP output, and moments for f-band elements were left out of the test set due to unreliable pseudopotentials. Models for predicting metal d-band centers, oxygen 2p-band centers, and magnetic moments were trained and tested on this bulk perovskite dataset using a multi-task learning scheme that learns these three properties simultaneously for every atom. Moreover, to train and test the model on Bader charges, over 170,000 structures with Bader charges available were compiled from across the entirety of Materials Project\cite{Jain2013}. This dataset includes a much more diverse set of structures than just perovskites. Lastly, to train and test the model on phonon band centers, a dataset of over 9,800 structures with phonon calculations was downloaded from the Phonon Database at Kyoto University\cite{PhononDatabase}. All structures available in this database were used, not limited to oxides and perovskites.

\begin{figure}
    \begin{center}
    \includegraphics[width=0.8\textwidth]{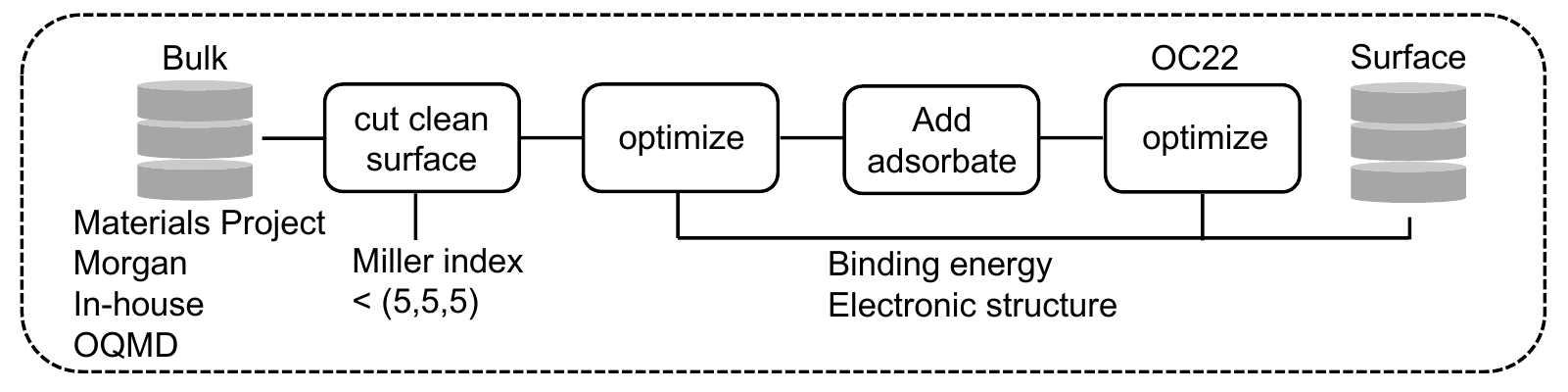}
    \end{center}
     \caption{In-house automation pipeline for generating the surfaces and binding energies dataset.}
     \label{surface_pipeline}
\end{figure}

We additionally prepared a novel dataset of perovskite oxide surfaces for predicting surface per-site properties and OER intermediate adsorption energies. This dataset is generated using an in-house automated pipeline (see Figure SI\ref{surface_pipeline}), where bulk perovskite structures up to quaternary are randomly sampled from the bulk perovskite dataset described above, cut with unique miller indices between (001) and (555), and optimized with and without adsorbed O$^*$, OH$^*$ and OOH$^*$. To account for the well-known overbinding of O$_2$, the oxygen reference was corrected to the experimental formation enthalpy of water, calculated as +0.29 V correction per O. Standard formation energies of H$_2$O(l), OH(g), and HO$_2$(g) were taken from NIST-JANAF thermochemical tables rather than being computed \cite{JANAF}. We combine this dataset with the data of clean surfaces in the Open Catalyst 2022 (OC22) dataset\cite{Tran2022} and surfaces with a single O$^*$, OH$^*$, and OOH$^*$ adsorbates totalling over 10,000 clean surfaces and 7,800 surfaces with adsorbates. All binding energies were converted to free energies by using a standard contribution of adsorbate atoms to the ZPE and vibrational entropy, calculated as +0.30, -0.01, and +0.34 V correction for O$^*$, OH$^*$ and OOH$^*$ respectively. Rotational and translational entropy of adsorbates on surfaces were assumed to be negligible, given that adsorption on surfaces typically results in loss of translational and rotational entropy in favor of vibrational entropy\cite{Norskov2014}.  Additional electronic structure calculations were performed on all structures from OC22 to generate per-site metal d-band center, O2p-band center and magnetic moment.

\begin{acknowledgement}

This work was supported by the Advanced Research Projects Agency–Energy (ARPA-E), US Department of Energy under award number DE-AR0001220. This research used resources of the National Energy Research Scientific Computing Center (NERSC), a U.S. Department of Energy Office of Science User Facility operated under Contract no. DE-AC02-05CH11231. We would additionally like to thank James Damewood for many fruitful discussions and suggestions, especially with regard to the bulk perovskite oxide data.

\end{acknowledgement}

\section{Author contributions}

Conceptualization, R.G.B., Y.S.H., J.R.L. and J.K.; Methodology, R.G.B., Y.S.H., J.R.L., J.K.; Formal analysis, J.R.L., J.K., and H.C.; Investigation, J.R.L., J.K., H.C., Y. T., and C.H.S.; Resources, R.G.B., Y.S.H. and B.H.; Writing - original draft, J.R.L., J.K., J.P., R.G.B., and Y.S.H.; Visualization, J.R.L. and J.K.; Supervision, R.G.B., Y.S.H., and B.H.; Funding Acquisition, R.G.B., Y.S.H., and B.H.

\section{Declaration of interests}

The authors declare no competing interests.


\bibliography{achemso-demo}

\providecommand{\latin}[1]{#1}
\makeatletter
\providecommand{\doi}
  {\begingroup\let\do\@makeother\dospecials
  \catcode`\{=1 \catcode`\}=2 \doi@aux}
\providecommand{\doi@aux}[1]{\endgroup\texttt{#1}}
\makeatother
\providecommand*\mcitethebibliography{\thebibliography}
\csname @ifundefined\endcsname{endmcitethebibliography}
  {\let\endmcitethebibliography\endthebibliography}{}
\begin{mcitethebibliography}{68}
\providecommand*\natexlab[1]{#1}
\providecommand*\mciteSetBstSublistMode[1]{}
\providecommand*\mciteSetBstMaxWidthForm[2]{}
\providecommand*\mciteBstWouldAddEndPuncttrue
  {\def\EndOfBibitem{\unskip.}}
\providecommand*\mciteBstWouldAddEndPunctfalse
  {\let\EndOfBibitem\relax}
\providecommand*\mciteSetBstMidEndSepPunct[3]{}
\providecommand*\mciteSetBstSublistLabelBeginEnd[3]{}
\providecommand*\EndOfBibitem{}
\mciteSetBstSublistMode{f}
\mciteSetBstMaxWidthForm{subitem}{(\alph{mcitesubitemcount})}
\mciteSetBstSublistLabelBeginEnd
  {\mcitemaxwidthsubitemform\space}
  {\relax}
  {\relax}

\bibitem[Hong \latin{et~al.}(2015)Hong, Risch, Stoerzinger, Grimaud,
  Suntivitch, and Shao-Horn]{Hong2015}
Hong,~W.~T.; Risch,~M.; Stoerzinger,~K.~A.; Grimaud,~A.; Suntivitch,~J.;
  Shao-Horn,~Y. Toward the rational design of non-precious transition metal
  oxides for oxygen electrocatalysis. \emph{Energy \& Environmental Science}
  \textbf{2015}, \emph{8}, 1404--1428\relax
\mciteBstWouldAddEndPuncttrue
\mciteSetBstMidEndSepPunct{\mcitedefaultmidpunct}
{\mcitedefaultendpunct}{\mcitedefaultseppunct}\relax
\EndOfBibitem
\bibitem[Dau \latin{et~al.}(2010)Dau, Limberg, Reier, Risch, Roggan, and
  Strasser]{Dau2010}
Dau,~H.; Limberg,~C.; Reier,~T.; Risch,~M.; Roggan,~S.; Strasser,~P. The
  Mechanism of Water Oxidation: From Electrolysis via Homogeneous to Biological
  Catalysis. \emph{ChemCatChem} \textbf{2010}, \emph{2}, 724--761\relax
\mciteBstWouldAddEndPuncttrue
\mciteSetBstMidEndSepPunct{\mcitedefaultmidpunct}
{\mcitedefaultendpunct}{\mcitedefaultseppunct}\relax
\EndOfBibitem
\bibitem[Cox \latin{et~al.}(2014)Cox, Retegan, Neese, Patnazis, Boussac, and
  Lubitz]{Cox2014}
Cox,~N.; Retegan,~M.; Neese,~F.; Patnazis,~D.~A.; Boussac,~A.; Lubitz,~W.
  Eectronic structure of the oxygen- evolving complex in photosystem II prior
  to O-O bond formation. \emph{Science} \textbf{2014}, \emph{345}\relax
\mciteBstWouldAddEndPuncttrue
\mciteSetBstMidEndSepPunct{\mcitedefaultmidpunct}
{\mcitedefaultendpunct}{\mcitedefaultseppunct}\relax
\EndOfBibitem
\bibitem[Vogt and Weckhuysen(2022)Vogt, and Weckhuysen]{Vogt2022}
Vogt,~C.; Weckhuysen,~B.~M. The concept of the active site in heterogeneous
  catalysis. \emph{Nature Reviews Chemistry} \textbf{2022}, \emph{6}\relax
\mciteBstWouldAddEndPuncttrue
\mciteSetBstMidEndSepPunct{\mcitedefaultmidpunct}
{\mcitedefaultendpunct}{\mcitedefaultseppunct}\relax
\EndOfBibitem
\bibitem[Norskov \latin{et~al.}(2014)Norskov, Studt, Abild-Pedersen, and
  Bligaard]{Norskov2014}
Norskov,~J.~K.; Studt,~F.; Abild-Pedersen,~F.; Bligaard,~T. Fundamental
  concepts in heterogeneous catalysis. \emph{John Wiley and Sons}
  \textbf{2014}, \emph{1}, 32\relax
\mciteBstWouldAddEndPuncttrue
\mciteSetBstMidEndSepPunct{\mcitedefaultmidpunct}
{\mcitedefaultendpunct}{\mcitedefaultseppunct}\relax
\EndOfBibitem
\bibitem[Ferri \latin{et~al.}(2023)Ferri, Li, Schwalbe-Koda, Xie, Moliner,
  G{\'o}mez-Bombarelli, Boronat, and Corma]{ferri2023approaching}
Ferri,~P.; Li,~C.; Schwalbe-Koda,~D.; Xie,~M.; Moliner,~M.;
  G{\'o}mez-Bombarelli,~R.; Boronat,~M.; Corma,~A. Approaching enzymatic
  catalysis with zeolites or how to select one reaction mechanism competing
  with others. \emph{Nature Communications} \textbf{2023}, \emph{14},
  2878\relax
\mciteBstWouldAddEndPuncttrue
\mciteSetBstMidEndSepPunct{\mcitedefaultmidpunct}
{\mcitedefaultendpunct}{\mcitedefaultseppunct}\relax
\EndOfBibitem
\bibitem[Rao \latin{et~al.}(2020)Rao, Kolb, Giordano, Pederson, Katayama,
  Hwang, Mehta, You, Lunger, Zhou, Halck, Vegge, Chorkendorff, Stephens, and
  Shao-Horn]{Rao2020}
Rao,~R.~R.; Kolb,~M.~J.; Giordano,~L.; Pederson,~A.~F.; Katayama,~Y.;
  Hwang,~J.; Mehta,~A.; You,~H.; Lunger,~J.~R.; Zhou,~H.; Halck,~N.~B.;
  Vegge,~T.; Chorkendorff,~I.; Stephens,~I.~E.; Shao-Horn,~Y. Operando
  identification of site-dependent water oxidation activity on ruthenium
  dioxide single-crystal surfaces. \emph{Nature Catalysis} \textbf{2020},
  \emph{3}, 516--525\relax
\mciteBstWouldAddEndPuncttrue
\mciteSetBstMidEndSepPunct{\mcitedefaultmidpunct}
{\mcitedefaultendpunct}{\mcitedefaultseppunct}\relax
\EndOfBibitem
\bibitem[Halck \latin{et~al.}(2014)Halck, Petrykin, Krtil, and
  Rossmeisl]{Halck2014}
Halck,~N.~B.; Petrykin,~V.; Krtil,~P.; Rossmeisl,~J. Beyond the Volcano
  Limitations in Electrocatalysis – Oxygen Evolution Reaction. \emph{Phys.
  Chem. Chem. Phys.} \textbf{2014}, \emph{16}, 13682--13688\relax
\mciteBstWouldAddEndPuncttrue
\mciteSetBstMidEndSepPunct{\mcitedefaultmidpunct}
{\mcitedefaultendpunct}{\mcitedefaultseppunct}\relax
\EndOfBibitem
\bibitem[Peng \latin{et~al.}(2022)Peng, Schwalbe-Koda, Akkiraju, Xie, Giordano,
  Yu, Eom, Lunger, Zheng, Rao, \latin{et~al.} others]{Peng2022}
Peng,~J.; Schwalbe-Koda,~D.; Akkiraju,~K.; Xie,~T.; Giordano,~L.; Yu,~Y.;
  Eom,~C.~J.; Lunger,~J.~R.; Zheng,~D.~J.; Rao,~R.~R., \latin{et~al.}
  Human-and machine-centred designs of molecules and materials for
  sustainability and decarbonization. \emph{Nature Reviews Materials}
  \textbf{2022}, 1--19\relax
\mciteBstWouldAddEndPuncttrue
\mciteSetBstMidEndSepPunct{\mcitedefaultmidpunct}
{\mcitedefaultendpunct}{\mcitedefaultseppunct}\relax
\EndOfBibitem
\bibitem[Hammer and Norskov(1995)Hammer, and Norskov]{Hammer1995}
Hammer,~B.; Norskov,~J.~K. Why gold is the noblest of all the metals.
  \emph{Nature} \textbf{1995}, \emph{376}, 238--240\relax
\mciteBstWouldAddEndPuncttrue
\mciteSetBstMidEndSepPunct{\mcitedefaultmidpunct}
{\mcitedefaultendpunct}{\mcitedefaultseppunct}\relax
\EndOfBibitem
\bibitem[Grimaud \latin{et~al.}(2013)Grimaud, May, Carlton, Lee, Risch, Hong,
  Zhuo, and Shao-Horn]{Grimaud2013}
Grimaud,~A.; May,~K.~J.; Carlton,~C.~E.; Lee,~Y.~L.; Risch,~M.; Hong,~W.~T.;
  Zhuo,~J.; Shao-Horn,~Y. Double perovskites as a family of highly active
  catalysts for oxygen evolution in alkaline solution. \emph{Nature
  Communications} \textbf{2013}, \emph{4}, 2439\relax
\mciteBstWouldAddEndPuncttrue
\mciteSetBstMidEndSepPunct{\mcitedefaultmidpunct}
{\mcitedefaultendpunct}{\mcitedefaultseppunct}\relax
\EndOfBibitem
\bibitem[Jacobs \latin{et~al.}(2018)Jacobs, Mayeshiba, Booske, and
  Morgan]{Jacobs2018}
Jacobs,~R.; Mayeshiba,~T.; Booske,~J.; Morgan,~D. Material Discovery and Design
  Principles for Stable, High Activity Perovskite Cathodes for Solid Oxide Fuel
  Cells. \emph{Advanced Energy Materials} \textbf{2018}, \emph{8},
  1702708\relax
\mciteBstWouldAddEndPuncttrue
\mciteSetBstMidEndSepPunct{\mcitedefaultmidpunct}
{\mcitedefaultendpunct}{\mcitedefaultseppunct}\relax
\EndOfBibitem
\bibitem[Suntivich \latin{et~al.}(2011)Suntivich, May, Gasteiger, Goodenough,
  and Shao-Horn]{Suntivich2011}
Suntivich,~J.; May,~K.~J.; Gasteiger,~H.~A.; Goodenough,~J.~B.; Shao-Horn,~Y. A
  Perovskite Oxide Optimized for Oxygen Evolution Catalysis from Molecular
  Orbital Principles. \emph{Science} \textbf{2011}, \emph{334},
  1383--1385\relax
\mciteBstWouldAddEndPuncttrue
\mciteSetBstMidEndSepPunct{\mcitedefaultmidpunct}
{\mcitedefaultendpunct}{\mcitedefaultseppunct}\relax
\EndOfBibitem
\bibitem[Biz \latin{et~al.}(2021)Biz, Fianchini, and Gracia]{Biz2021}
Biz,~C.; Fianchini,~M.; Gracia,~J. Strongly Correlated Electrons in Catalysis:
  Focus on Quantum Exchange. \emph{ACS Catalysis} \textbf{2021}, \emph{11},
  14249--14261\relax
\mciteBstWouldAddEndPuncttrue
\mciteSetBstMidEndSepPunct{\mcitedefaultmidpunct}
{\mcitedefaultendpunct}{\mcitedefaultseppunct}\relax
\EndOfBibitem
\bibitem[Vennelakanti \latin{et~al.}(2022)Vennelakanti, Nandy, and
  Kulik]{Vennelakanti2022}
Vennelakanti,~V.; Nandy,~A.; Kulik,~H.~J. The Effect of Hartree‑Fock Exchange
  on Scaling Relations and Reaction Energetics for C–H Activation Catalysts.
  \emph{Topics in Catalysis} \textbf{2022}, \emph{65}, 296–311\relax
\mciteBstWouldAddEndPuncttrue
\mciteSetBstMidEndSepPunct{\mcitedefaultmidpunct}
{\mcitedefaultendpunct}{\mcitedefaultseppunct}\relax
\EndOfBibitem
\bibitem[Grimaud \latin{et~al.}(2017)Grimaud, Diaz-Morales, Han, Hong, Lee,
  Giordano, Stoerzinger, Koper, and Shao-Horn]{Grimaud2017}
Grimaud,~A.; Diaz-Morales,~O.; Han,~B.; Hong,~W.~T.; Lee,~Y.~L.; Giordano,~L.;
  Stoerzinger,~K.~A.; Koper,~M. T.~M.; Shao-Horn,~Y. Activating lattice oxygen
  redox reactions in metal oxides to catalyse oxygen evolution. \emph{Nature
  Chemistry} \textbf{2017}, \emph{9}, 457--465\relax
\mciteBstWouldAddEndPuncttrue
\mciteSetBstMidEndSepPunct{\mcitedefaultmidpunct}
{\mcitedefaultendpunct}{\mcitedefaultseppunct}\relax
\EndOfBibitem
\bibitem[Kuznetsov \latin{et~al.}(2020)Kuznetsov, Peng, Giordano,
  Román-Leshkov, and Shao-Horn]{Kuznetsov2020}
Kuznetsov,~D.~A.; Peng,~J.; Giordano,~L.; Román-Leshkov,~Y.; Shao-Horn,~Y.
  Bismuth Substituted Strontium Cobalt Perovskites for Catalyzing Oxygen
  Evolution. \emph{J. Phys. Chem. C} \textbf{2020}, \emph{124},
  6562--6570\relax
\mciteBstWouldAddEndPuncttrue
\mciteSetBstMidEndSepPunct{\mcitedefaultmidpunct}
{\mcitedefaultendpunct}{\mcitedefaultseppunct}\relax
\EndOfBibitem
\bibitem[Yuan \latin{et~al.}(2022)Yuan, Peng, Cai, Huang, Garcia-Esparza,
  Sokaras, Zhang, Giordano, Akkiraju, Zhu, Hubner, Zou, Roman-Leshkov, and
  Shao-Horn]{Yuan2022}
Yuan,~S.; Peng,~J.; Cai,~B.; Huang,~Z.; Garcia-Esparza,~A.~T.; Sokaras,~D.;
  Zhang,~Y.; Giordano,~L.; Akkiraju,~K.; Zhu,~Y.~G.; Hubner,~R.; Zou,~X.;
  Roman-Leshkov,~Y.; Shao-Horn,~Y. Tunable metal hydroxide–organic frameworks
  for catalysing oxygen evolution. \emph{Nature Materials} \textbf{2022},
  \emph{21}, 673--680\relax
\mciteBstWouldAddEndPuncttrue
\mciteSetBstMidEndSepPunct{\mcitedefaultmidpunct}
{\mcitedefaultendpunct}{\mcitedefaultseppunct}\relax
\EndOfBibitem
\bibitem[Calle-Vallejo \latin{et~al.}(2013)Calle-Vallejo, Inoglu, Su, Kitchin,
  and Rossmeisl]{CalleVallejo2013}
Calle-Vallejo,~F.; Inoglu,~N.~G.; Su,~H.~Y.; Kitchin,~J.~R.; Rossmeisl,~J.
  Number of outer electrons as descriptor for adsorption processes on
  transition metals and their oxides. \emph{Chemical Science} \textbf{2013},
  \emph{4}, 1245--1249\relax
\mciteBstWouldAddEndPuncttrue
\mciteSetBstMidEndSepPunct{\mcitedefaultmidpunct}
{\mcitedefaultendpunct}{\mcitedefaultseppunct}\relax
\EndOfBibitem
\bibitem[Dickens and Norksov(2017)Dickens, and Norksov]{Dickens2017}
Dickens,~C.~F.; Norksov,~J.~K. A Theoretical Investigation into the Role of
  Surface Defects for Oxygen Evolution on RuO2. \emph{J. Phys. Chem. C.}
  \textbf{2017}, \emph{121}, 18516–18524\relax
\mciteBstWouldAddEndPuncttrue
\mciteSetBstMidEndSepPunct{\mcitedefaultmidpunct}
{\mcitedefaultendpunct}{\mcitedefaultseppunct}\relax
\EndOfBibitem
\bibitem[Choubisa \latin{et~al.}(2023)Choubisa, Abed, Mendoza, Sutherland,
  Apuru-Guzik, and Sargent]{Choubisa2023}
Choubisa,~H.; Abed,~J.; Mendoza,~D.; Sutherland,~B.~R.; Apuru-Guzik,~A.;
  Sargent,~E.~H. Accelerated chemical space search using a quantum-inspired
  cluster expansion approach. \emph{Matter} \textbf{2023}, \emph{6}\relax
\mciteBstWouldAddEndPuncttrue
\mciteSetBstMidEndSepPunct{\mcitedefaultmidpunct}
{\mcitedefaultendpunct}{\mcitedefaultseppunct}\relax
\EndOfBibitem
\bibitem[Rao \latin{et~al.}(2017)Rao, Kolb, Halck, Pedersen, Mehta, You,
  Stoerzinger, Feng, Hansen, Zhou, Giordano, Rossmeisl, Vegge, Chorkendorff,
  Stephens, and Shao-Horn]{Rao2017}
Rao,~R.~R. \latin{et~al.}  Towards identifying the active sites on RuO2(110) in
  catalyzing oxygen evolution. \emph{Energy Eviron. Sci.} \textbf{2017},
  \emph{10}, 2626--2637\relax
\mciteBstWouldAddEndPuncttrue
\mciteSetBstMidEndSepPunct{\mcitedefaultmidpunct}
{\mcitedefaultendpunct}{\mcitedefaultseppunct}\relax
\EndOfBibitem
\bibitem[Dickens \latin{et~al.}(2019)Dickens, Montoya, Kulkarni, Bajdich, and
  Norksov]{Dickens2019}
Dickens,~C.; Montoya,~J.~H.; Kulkarni,~A.~R.; Bajdich,~M.; Norksov,~J.~K. An
  electronic structure descriptor for oxygen reactivity at metal and metal-
  oxide surfaces. \emph{Surface Science} \textbf{2019}, \emph{681}\relax
\mciteBstWouldAddEndPuncttrue
\mciteSetBstMidEndSepPunct{\mcitedefaultmidpunct}
{\mcitedefaultendpunct}{\mcitedefaultseppunct}\relax
\EndOfBibitem
\bibitem[Hwang \latin{et~al.}(2021)Hwang, Rao, Giordano, Akkiraju, Wang,
  Crumlin, Bluhm, and Shao-Horn]{Hwang2021}
Hwang,~J.; Rao,~R.~R.; Giordano,~L.; Akkiraju,~K.; Wang,~X.~R.; Crumlin,~E.~J.;
  Bluhm,~H.; Shao-Horn,~Y. Regulating oxygen activity of perovskites to promote
  NOx oxidation and reduction kinetics. \emph{Nature Catalysis} \textbf{2021},
  \emph{4}, 663--673\relax
\mciteBstWouldAddEndPuncttrue
\mciteSetBstMidEndSepPunct{\mcitedefaultmidpunct}
{\mcitedefaultendpunct}{\mcitedefaultseppunct}\relax
\EndOfBibitem
\bibitem[Calle-Vallejo \latin{et~al.}(2015)Calle-Vallejo, Loffreda, Koper, and
  Sautet]{CalleVallejo2015}
Calle-Vallejo,~F.; Loffreda,~D.; Koper,~M. T.~M.; Sautet,~P. Introducing
  structural sensitivity into adsorption–energy scaling relations by means of
  coordination numbers. \emph{Nature Chemistry} \textbf{2015}, \emph{7},
  403--410\relax
\mciteBstWouldAddEndPuncttrue
\mciteSetBstMidEndSepPunct{\mcitedefaultmidpunct}
{\mcitedefaultendpunct}{\mcitedefaultseppunct}\relax
\EndOfBibitem
\bibitem[Ruck \latin{et~al.}(2018)Ruck, Bandarenka, Calle-Vallejo, and
  Gagliardi]{Ruck2018}
Ruck,~M.; Bandarenka,~A.; Calle-Vallejo,~F.; Gagliardi,~A. Oxygen Reduction
  Reaction: Rapid Prediction of Mass Activity of Nanostructured Platinum
  Electrocatalysts. \emph{J. Phys. Chem. Lett.} \textbf{2018}, \emph{9},
  4463--4468\relax
\mciteBstWouldAddEndPuncttrue
\mciteSetBstMidEndSepPunct{\mcitedefaultmidpunct}
{\mcitedefaultendpunct}{\mcitedefaultseppunct}\relax
\EndOfBibitem
\bibitem[Batchelor \latin{et~al.}(2019)Batchelor, Pederson, Winther, Castelli,
  Jacobsen, and Rossmeisl]{Batchelor2019}
Batchelor,~T. A.~A.; Pederson,~J.~K.; Winther,~S.; Castelli,~I.~E.;
  Jacobsen,~K.~W.; Rossmeisl,~J. High-Entropy Alloys as a Discovery Platform
  for Electrocatalysis. \emph{Joule} \textbf{2019}, \emph{3}, 834--845\relax
\mciteBstWouldAddEndPuncttrue
\mciteSetBstMidEndSepPunct{\mcitedefaultmidpunct}
{\mcitedefaultendpunct}{\mcitedefaultseppunct}\relax
\EndOfBibitem
\bibitem[Tran and Ulissi(2011)Tran, and Ulissi]{Tran2018}
Tran,~K.; Ulissi,~Z.~W. Active learning across intermetallics to guide
  discovery of electrocatalysts for CO2 reduction and H2 evolution.
  \emph{Nature Catalysis} \textbf{2011}, \emph{1}, 696--703\relax
\mciteBstWouldAddEndPuncttrue
\mciteSetBstMidEndSepPunct{\mcitedefaultmidpunct}
{\mcitedefaultendpunct}{\mcitedefaultseppunct}\relax
\EndOfBibitem
\bibitem[Fernandez \latin{et~al.}(2017)Fernandez, Barron, and
  S.~Barnard]{Fernandez2017}
Fernandez,~M.; Barron,~H.; S.~Barnard,~A. {Artificial neural network analysis
  of the catalytic efficiency of platinum nanoparticles}. \emph{RSC Adv.}
  \textbf{2017}, \emph{7}, 48962\relax
\mciteBstWouldAddEndPuncttrue
\mciteSetBstMidEndSepPunct{\mcitedefaultmidpunct}
{\mcitedefaultendpunct}{\mcitedefaultseppunct}\relax
\EndOfBibitem
\bibitem[Li \latin{et~al.}(2017)Li, Ma, and Xin]{Li2017}
Li,~Z.; Ma,~X.; Xin,~H. {Feature engineering of machine-learning chemisorption
  models for catalyst design}. \emph{Catalysis Today} \textbf{2017},
  \emph{280}, 232--238\relax
\mciteBstWouldAddEndPuncttrue
\mciteSetBstMidEndSepPunct{\mcitedefaultmidpunct}
{\mcitedefaultendpunct}{\mcitedefaultseppunct}\relax
\EndOfBibitem
\bibitem[Axelrod \latin{et~al.}(2022)Axelrod, Schwalbe-Koda, Mohapatra,
  Damewood, Greenman, and G{\'o}mez-Bombarelli]{Axelrod2022}
Axelrod,~S.; Schwalbe-Koda,~D.; Mohapatra,~S.; Damewood,~J.; Greenman,~K.~P.;
  G{\'o}mez-Bombarelli,~R. Learning matter: Materials design with machine
  learning and atomistic simulations. \emph{Accounts of Materials Research}
  \textbf{2022}, \emph{3}, 343--357\relax
\mciteBstWouldAddEndPuncttrue
\mciteSetBstMidEndSepPunct{\mcitedefaultmidpunct}
{\mcitedefaultendpunct}{\mcitedefaultseppunct}\relax
\EndOfBibitem
\bibitem[Kitchin(2018)]{Kitchin2018}
Kitchin,~J.~R. Machine learning in catalysis. \emph{Nature Catalysis}
  \textbf{2018}, \emph{1}, 230--232\relax
\mciteBstWouldAddEndPuncttrue
\mciteSetBstMidEndSepPunct{\mcitedefaultmidpunct}
{\mcitedefaultendpunct}{\mcitedefaultseppunct}\relax
\EndOfBibitem
\bibitem[Xie and Grossman(2018)Xie, and Grossman]{Xie2018}
Xie,~T.; Grossman,~J.~C. Crystal Graph Convolutional Neural Networks for an
  Accurate and Interpretable Prediction of Material Properties. \emph{Physical
  Review Letters} \textbf{2018}, \emph{120}, 145301\relax
\mciteBstWouldAddEndPuncttrue
\mciteSetBstMidEndSepPunct{\mcitedefaultmidpunct}
{\mcitedefaultendpunct}{\mcitedefaultseppunct}\relax
\EndOfBibitem
\bibitem[Damewood \latin{et~al.}(2023)Damewood, Karaguesian, Lunger, Tan, Xie,
  Peng, and G{\'o}mez-Bombarelli]{Damewood2023}
Damewood,~J.; Karaguesian,~J.; Lunger,~J.~R.; Tan,~A.~R.; Xie,~M.; Peng,~J.;
  G{\'o}mez-Bombarelli,~R. Representations of Materials for Machine Learning.
  \emph{Annual Review of Materials Research} \textbf{2023}, \emph{53}\relax
\mciteBstWouldAddEndPuncttrue
\mciteSetBstMidEndSepPunct{\mcitedefaultmidpunct}
{\mcitedefaultendpunct}{\mcitedefaultseppunct}\relax
\EndOfBibitem
\bibitem[Greenman \latin{et~al.}(2022)Greenman, Green, and
  G{\'o}mez-Bombarelli]{Greenman2022}
Greenman,~K.~P.; Green,~W.~H.; G{\'o}mez-Bombarelli,~R. Multi-fidelity
  prediction of molecular optical peaks with deep learning. \emph{Chem. Sci.}
  \textbf{2022}, \emph{13}, 1152--1162\relax
\mciteBstWouldAddEndPuncttrue
\mciteSetBstMidEndSepPunct{\mcitedefaultmidpunct}
{\mcitedefaultendpunct}{\mcitedefaultseppunct}\relax
\EndOfBibitem
\bibitem[Tran \latin{et~al.}(2022)Tran, Lan, Shuaibi, Wood, Goyal, Das,
  Heras-Domingo, Kolluru, Rizvi, Shoghi, Sriram, Therrien, Abed, Voznyy,
  Sargent, Ulissi, and Zitnick]{Tran2022}
Tran,~R. \latin{et~al.}  The Open Catalyst 2022 (OC22) Dataset and Challenges
  for Oxide Electrocatalysts. \emph{arXiv} \textbf{2022}, \relax
\mciteBstWouldAddEndPunctfalse
\mciteSetBstMidEndSepPunct{\mcitedefaultmidpunct}
{}{\mcitedefaultseppunct}\relax
\EndOfBibitem
\bibitem[Back \latin{et~al.}(2019)Back, Yoon, Tian, Zhong, Tran, and
  Ulissi]{Back2019}
Back,~S.; Yoon,~J.; Tian,~N.; Zhong,~W.; Tran,~K.; Ulissi,~Z.~W. Convolutional
  Neural Network of Atomic Surface Structures To Predict Binding Energies for
  High-Throughput Screening of Catalysts. \emph{J. Phys. Chem. Lett.}
  \textbf{2019}, \emph{10}, 4401--4408\relax
\mciteBstWouldAddEndPuncttrue
\mciteSetBstMidEndSepPunct{\mcitedefaultmidpunct}
{\mcitedefaultendpunct}{\mcitedefaultseppunct}\relax
\EndOfBibitem
\bibitem[Hwang \latin{et~al.}(2017)Hwang, Rao, Giordano, Katayama, Yu, and
  Shao-Horn]{Hwang2017}
Hwang,~J.; Rao,~R.~R.; Giordano,~L.; Katayama,~Y.; Yu,~Y.; Shao-Horn,~Y.
  Perovskites in catalysis and electrocatalysis. \emph{Science} \textbf{2017},
  \emph{358}, 751–756\relax
\mciteBstWouldAddEndPuncttrue
\mciteSetBstMidEndSepPunct{\mcitedefaultmidpunct}
{\mcitedefaultendpunct}{\mcitedefaultseppunct}\relax
\EndOfBibitem
\bibitem[Rawal \latin{et~al.}(2019)Rawal, Ozcan, Liu, Pingali, Akbilgic,
  Tetard, O'Neill, Santra, and Petridis]{Rawal2019}
Rawal,~T.~B.; Ozcan,~A.; Liu,~S.~H.; Pingali,~S.~V.; Akbilgic,~O.; Tetard,~L.;
  O'Neill,~H.; Santra,~S.; Petridis,~L. {Interaction of Zinc Oxide
  Nanoparticles with Water: Implications for Catalytic Activity}. \emph{ACS
  Appl. Nano Mater.} \textbf{2019}, \emph{2}, 4257--4266\relax
\mciteBstWouldAddEndPuncttrue
\mciteSetBstMidEndSepPunct{\mcitedefaultmidpunct}
{\mcitedefaultendpunct}{\mcitedefaultseppunct}\relax
\EndOfBibitem
\bibitem[Zhong \latin{et~al.}(2021)Zhong, Qiu, Shen, Wang, Yuan, Jia, Bi, and
  Jiang]{Zhong2021}
Zhong,~W.; Qiu,~Y.; Shen,~H.; Wang,~X.; Yuan,~J.; Jia,~C.; Bi,~S.; Jiang,~J.
  {Electronic Spin Moment As a Catalytic Descriptor for Fe Single-Atom
  Catalysts Supported on C2N}. \emph{J. Am. Chem. Soc.} \textbf{2021},
  \emph{143}, 4405--4413\relax
\mciteBstWouldAddEndPuncttrue
\mciteSetBstMidEndSepPunct{\mcitedefaultmidpunct}
{\mcitedefaultendpunct}{\mcitedefaultseppunct}\relax
\EndOfBibitem
\bibitem[Castelli \latin{et~al.}(2012)Castelli, Olsen, Dutta, Landis, Dahl,
  Thygesen, and Jacobsen]{Castelli2012}
Castelli,~I.~E.; Olsen,~T.; Dutta,~S.; Landis,~D.~D.; Dahl,~S.;
  Thygesen,~K.~S.; Jacobsen,~K.~W. Computational screening of perovskite metal
  oxides for optimal solar light capture. \emph{Energy \& Environmental
  Science} \textbf{2012}, \emph{5}, 5814--5819\relax
\mciteBstWouldAddEndPuncttrue
\mciteSetBstMidEndSepPunct{\mcitedefaultmidpunct}
{\mcitedefaultendpunct}{\mcitedefaultseppunct}\relax
\EndOfBibitem
\bibitem[Emery \latin{et~al.}(2016)Emery, Saal, Kirklin, Hegde, and
  Wolverton]{Emery2016}
Emery,~A.~A.; Saal,~J.~E.; Kirklin,~S.; Hegde,~V.~I.; Wolverton,~C.
  High-Throughput Computational Screening of Perovskites for Thermochemical
  Water Splitting Applications. \emph{Chemistry of Materials} \textbf{2016},
  \emph{28}, 5621–5634\relax
\mciteBstWouldAddEndPuncttrue
\mciteSetBstMidEndSepPunct{\mcitedefaultmidpunct}
{\mcitedefaultendpunct}{\mcitedefaultseppunct}\relax
\EndOfBibitem
\bibitem[Jain \latin{et~al.}(2013)Jain, Ong, Hauttier, Chen, Richards, Dacek,
  Cholia, Gunter, Skinner, Ceder, and Persson]{Jain2013}
Jain,~A.; Ong,~S.~P.; Hauttier,~G.; Chen,~W.; Richards,~W.~D.; Dacek,~S.;
  Cholia,~S.; Gunter,~D.; Skinner,~D.; Ceder,~G.; Persson,~K.~A. Commentary:
  The Materials Project: A materials genome approach to accelerating materials
  innovation. \emph{APL Materials} \textbf{2013}, \emph{1}, 011002\relax
\mciteBstWouldAddEndPuncttrue
\mciteSetBstMidEndSepPunct{\mcitedefaultmidpunct}
{\mcitedefaultendpunct}{\mcitedefaultseppunct}\relax
\EndOfBibitem
\bibitem[Kulkarni \latin{et~al.}(2018)Kulkarni, Siahrostami, Patel, and
  Norskov]{Kulkarni2018}
Kulkarni,~A.; Siahrostami,~S.; Patel,~A.; Norskov,~J.~K. Understanding
  Catalytic Activity Trends in the Oxygen Reduction Reaction. \emph{Chem. Rev.}
  \textbf{2018}, \emph{5}, 2302--2312\relax
\mciteBstWouldAddEndPuncttrue
\mciteSetBstMidEndSepPunct{\mcitedefaultmidpunct}
{\mcitedefaultendpunct}{\mcitedefaultseppunct}\relax
\EndOfBibitem
\bibitem[Gauthier \latin{et~al.}(2017)Gauthier, Dickens, Chen, Doyle, and
  Norskov]{Gauthier2017}
Gauthier,~J.~A.; Dickens,~C.~F.; Chen,~L.~D.; Doyle,~A.~D.; Norskov,~J.~K.
  Solvation Effects for Oxygen Evolution Reaction Catalysis on IrO2(110).
  \emph{J. Phys. Chem. C.} \textbf{2017}, \emph{121}, 11455--11463\relax
\mciteBstWouldAddEndPuncttrue
\mciteSetBstMidEndSepPunct{\mcitedefaultmidpunct}
{\mcitedefaultendpunct}{\mcitedefaultseppunct}\relax
\EndOfBibitem
\bibitem[Deng \latin{et~al.}(2019)Deng, Zhao, Wu, Chen, Hansen, and
  Vegge]{Deng2019}
Deng,~Q.; Zhao,~J.; Wu,~T.; Chen,~G.; Hansen,~H.~A.; Vegge,~T. 2D transition
  metal–TCNQ sheets as bifunctional single-atom catalysts. \emph{Journal of
  Catalysis} \textbf{2019}, \emph{370}, 378–384\relax
\mciteBstWouldAddEndPuncttrue
\mciteSetBstMidEndSepPunct{\mcitedefaultmidpunct}
{\mcitedefaultendpunct}{\mcitedefaultseppunct}\relax
\EndOfBibitem
\bibitem[Siegbahn(2013)]{Siegbahn2013}
Siegbahn,~P. E.~M. Water oxidation mechanism in photosystem II, including
  oxidations, proton release pathways, O―O bond formation and O2 release.
  \emph{Biochimica et Biophysica Acta} \textbf{2013}, \emph{1827},
  1003--1019\relax
\mciteBstWouldAddEndPuncttrue
\mciteSetBstMidEndSepPunct{\mcitedefaultmidpunct}
{\mcitedefaultendpunct}{\mcitedefaultseppunct}\relax
\EndOfBibitem
\bibitem[Pho(2022)]{PhononDatabase}
Phonon database at Kyoto University -- phonondb documentation.
  \emph{Phonondb.mtl.kyoto-u.ac.jp [Online]} \textbf{2022}, \relax
\mciteBstWouldAddEndPunctfalse
\mciteSetBstMidEndSepPunct{\mcitedefaultmidpunct}
{}{\mcitedefaultseppunct}\relax
\EndOfBibitem
\bibitem[Lohmiller \latin{et~al.}(2014)Lohmiller, Krewald, Navarro, Retegan,
  Rapatskiy, Nowaczyk, Boussac, Neese, Lubitz, Pantazis, and
  Cox]{Lohmiller2014}
Lohmiller,~T.; Krewald,~V.; Navarro,~M.~P.; Retegan,~M.; Rapatskiy,~L.;
  Nowaczyk,~M.; Boussac,~A.; Neese,~F.; Lubitz,~W.; Pantazis,~D.~A.; Cox,~N.
  Structure, ligands and substrate coordination of the oxygen-evolving complex
  of photosystem II in the S2 state: a combined EPR and DFT study. \emph{Phys.
  Chem. Chem. Phys.} \textbf{2014}, \emph{16}, 11877--11892\relax
\mciteBstWouldAddEndPuncttrue
\mciteSetBstMidEndSepPunct{\mcitedefaultmidpunct}
{\mcitedefaultendpunct}{\mcitedefaultseppunct}\relax
\EndOfBibitem
\bibitem[Schutt \latin{et~al.}(2021)Schutt, Unke, and Gastegger]{Schutt2021}
Schutt,~K.~T.; Unke,~O.~T.; Gastegger,~M. Equivariant message passing for the
  prediction of tensorial properties and molecular spectra. \emph{arXiv}
  \textbf{2021}, \relax
\mciteBstWouldAddEndPunctfalse
\mciteSetBstMidEndSepPunct{\mcitedefaultmidpunct}
{}{\mcitedefaultseppunct}\relax
\EndOfBibitem
\bibitem[O'Keefe and Breese(1991)O'Keefe, and Breese]{OKeefe1991}
O'Keefe,~M.; Breese,~N.~E. Atom sizes and bond lengths in molecules and
  crystals. \emph{J. Am. Chem. Soc.} \textbf{1991}, \emph{113},
  3226–3229\relax
\mciteBstWouldAddEndPuncttrue
\mciteSetBstMidEndSepPunct{\mcitedefaultmidpunct}
{\mcitedefaultendpunct}{\mcitedefaultseppunct}\relax
\EndOfBibitem
\bibitem[Kutzelnigg and Morgan(1996)Kutzelnigg, and Morgan]{Kutzelnigg1996}
Kutzelnigg,~W.; Morgan,~J. D.~I. {Hund's rules}. \emph{Z. Phys. D.}
  \textbf{1996}, \emph{36}, 197--214\relax
\mciteBstWouldAddEndPuncttrue
\mciteSetBstMidEndSepPunct{\mcitedefaultmidpunct}
{\mcitedefaultendpunct}{\mcitedefaultseppunct}\relax
\EndOfBibitem
\bibitem[Jacobs \latin{et~al.}(2019)Jacobs, Hwang, Shao-Horn, and
  Morgan]{Jacobs2019}
Jacobs,~R.; Hwang,~J.; Shao-Horn,~Y.; Morgan,~D. Assessing Correlations of
  Perovskite Catalytic Performance with Electronic Structure Descriptors.
  \emph{Chemistry of Materials} \textbf{2019}, \emph{3}, 785--797\relax
\mciteBstWouldAddEndPuncttrue
\mciteSetBstMidEndSepPunct{\mcitedefaultmidpunct}
{\mcitedefaultendpunct}{\mcitedefaultseppunct}\relax
\EndOfBibitem
\bibitem[Chen and Ong(2022)Chen, and Ong]{Chen2022}
Chen,~C.; Ong,~S. A universal graph deep learning interatomic potential for the
  periodic table. \emph{Nature Computational Science} \textbf{2022},
  718--728\relax
\mciteBstWouldAddEndPuncttrue
\mciteSetBstMidEndSepPunct{\mcitedefaultmidpunct}
{\mcitedefaultendpunct}{\mcitedefaultseppunct}\relax
\EndOfBibitem
\bibitem[Deng \latin{et~al.}(2023)Deng, Zhong, Jun, Han, Bartel, and
  Ceder]{Deng2023}
Deng,~B.; Zhong,~P.; Jun,~K.; Han,~K.; Bartel,~C.~J.; Ceder,~B. CHGNet:
  Pretrained universal neural network potential for charge-informed atomistic
  modeling. \emph{arXiv} \textbf{2023}, \relax
\mciteBstWouldAddEndPunctfalse
\mciteSetBstMidEndSepPunct{\mcitedefaultmidpunct}
{}{\mcitedefaultseppunct}\relax
\EndOfBibitem
\bibitem[Lee \latin{et~al.}(2011)Lee, Kleis, Rossmeisl, Shao-Horn, and
  Morgan]{Lee2011}
Lee,~Y.~L.; Kleis,~J.; Rossmeisl,~J.; Shao-Horn,~Y.; Morgan,~D. Prediction of
  solid oxide fuel cell cathode activity with first-principles descriptors.
  \emph{Energy and Environmental Science} \textbf{2011}, \emph{4}, 3966\relax
\mciteBstWouldAddEndPuncttrue
\mciteSetBstMidEndSepPunct{\mcitedefaultmidpunct}
{\mcitedefaultendpunct}{\mcitedefaultseppunct}\relax
\EndOfBibitem
\bibitem[Craig \latin{et~al.}(2023)Craig, Kleuker, Badjdich, and
  Garcia-Melchor]{Craig2023}
Craig,~M.~J.; Kleuker,~F.; Badjdich,~M.; Garcia-Melchor,~M. FEFOS: A method to
  derive oxide formation energies from oxidation states. \emph{ChemRxiv}
  \textbf{2023}, \relax
\mciteBstWouldAddEndPunctfalse
\mciteSetBstMidEndSepPunct{\mcitedefaultmidpunct}
{}{\mcitedefaultseppunct}\relax
\EndOfBibitem
\bibitem[Hong \latin{et~al.}(2016)Hong, Welsch, and Shao-Horn]{Hong2016}
Hong,~W.~T.; Welsch,~R.~E.; Shao-Horn,~Y. Descriptors of Oxygen-Evolution
  Activity for Oxides: A Statistical Evaluation. \emph{J. Phys. Chem. C.}
  \textbf{2016}, \emph{120}, 78--86\relax
\mciteBstWouldAddEndPuncttrue
\mciteSetBstMidEndSepPunct{\mcitedefaultmidpunct}
{\mcitedefaultendpunct}{\mcitedefaultseppunct}\relax
\EndOfBibitem
\bibitem[Sun \latin{et~al.}(2020)Sun, Zhou, Cong, Hong, and Chen]{Sun2020}
Sun,~S.; Zhou,~X.; Cong,~B.; Hong,~W.; Chen,~G. Tailoring the d-Band Centers
  Endows (NixFe1–x)2P Nanosheets with Efficient Oxygen Evolution Catalysis.
  \emph{ACS Catal.} \textbf{2020}, \emph{10}, 9086–9097\relax
\mciteBstWouldAddEndPuncttrue
\mciteSetBstMidEndSepPunct{\mcitedefaultmidpunct}
{\mcitedefaultendpunct}{\mcitedefaultseppunct}\relax
\EndOfBibitem
\bibitem[Man \latin{et~al.}(2011)Man, Su, Calle-Vallejo, Hansen, Martinez,
  Inoglu, Kitchin, Jaramillo, Norksov, and Rossmeisl]{Man2011}
Man,~I.~C.; Su,~H.; Calle-Vallejo,~F.; Hansen,~H.~A.; Martinez,~J.~I.;
  Inoglu,~N.~G.; Kitchin,~J.; Jaramillo,~T.~F.; Norksov,~J.~K.; Rossmeisl,~J.
  Universality in Oxygen Evolution Electrocatalysis on Oxide Surfaces.
  \emph{ChemCatChem} \textbf{2011}, \emph{3}, 1159–1165\relax
\mciteBstWouldAddEndPuncttrue
\mciteSetBstMidEndSepPunct{\mcitedefaultmidpunct}
{\mcitedefaultendpunct}{\mcitedefaultseppunct}\relax
\EndOfBibitem
\bibitem[Rossmeisl \latin{et~al.}(2007)Rossmeisl, Dimitrievski, Siegbahn, and
  Norskov]{Rossmeisl2007}
Rossmeisl,~J.; Dimitrievski,~K.; Siegbahn,~P.; Norskov,~J.~K. Comparing
  Electrochemical and Biological Water Splitting. \emph{The Journal of Physical
  Chemistry Letters C.} \textbf{2007}, \emph{111}, 18821--18823\relax
\mciteBstWouldAddEndPuncttrue
\mciteSetBstMidEndSepPunct{\mcitedefaultmidpunct}
{\mcitedefaultendpunct}{\mcitedefaultseppunct}\relax
\EndOfBibitem
\bibitem[Pedersen \latin{et~al.}(2021)Pedersen, Clausen, Krysiak, Xiao,
  Batchelor, Loffler, Mints, Banko, Arenz, Savan, Schumann, Ludwig, and
  Rossmeisl]{Pedersen2019}
Pedersen,~J.~K.; Clausen,~C.~M.; Krysiak,~O.~A.; Xiao,~B.; Batchelor,~T. A.~A.;
  Loffler,~T.; Mints,~V.~A.; Banko,~L.; Arenz,~M.; Savan,~A.; Schumann,~W.;
  Ludwig,~A.; Rossmeisl,~J. Bayesian Optimization of High-Entropy Alloy
  Compositions for Electrocatalytic Oxygen Reduction. \emph{Angewandte Chemie}
  \textbf{2021}, \emph{1333}, 24346--24354\relax
\mciteBstWouldAddEndPuncttrue
\mciteSetBstMidEndSepPunct{\mcitedefaultmidpunct}
{\mcitedefaultendpunct}{\mcitedefaultseppunct}\relax
\EndOfBibitem
\bibitem[Svane and Rossmeisl(2022)Svane, and Rossmeisl]{Svane2022}
Svane,~K.; Rossmeisl,~J. High-entropy oxides for the oxygen evolution reaction.
  \emph{ChemRxiv. Cambridge Open Engage} \textbf{2022}, \emph{1}\relax
\mciteBstWouldAddEndPuncttrue
\mciteSetBstMidEndSepPunct{\mcitedefaultmidpunct}
{\mcitedefaultendpunct}{\mcitedefaultseppunct}\relax
\EndOfBibitem
\bibitem[Nguyen \latin{et~al.}(2021)Nguyen, Liao, Lin, Su, and
  Ting]{Nguyen2021}
Nguyen,~T.~X.; Liao,~Y.~C.; Lin,~C.~C.; Su,~Y.~H.; Ting,~J.~M. Advanced High
  Entropy Perovskite Oxide Electrocatalyst for Oxygen Evolution Reaction.
  \emph{Adv. Funct. Mater.} \textbf{2021}, \emph{31}, 2101632\relax
\mciteBstWouldAddEndPuncttrue
\mciteSetBstMidEndSepPunct{\mcitedefaultmidpunct}
{\mcitedefaultendpunct}{\mcitedefaultseppunct}\relax
\EndOfBibitem
\bibitem[Peng \latin{et~al.}(2023)Peng, Damewood, and
  G{\'o}mez-Bombarelli]{Peng2023}
Peng,~J.; Damewood,~J.; G{\'o}mez-Bombarelli,~R. Data-Driven, Physics-Informed
  Descriptors of Cation Ordering in Multicomponent Oxides. \emph{arXiv}
  \textbf{2023}, \relax
\mciteBstWouldAddEndPunctfalse
\mciteSetBstMidEndSepPunct{\mcitedefaultmidpunct}
{}{\mcitedefaultseppunct}\relax
\EndOfBibitem
\bibitem[Clark and Hayes(2019)Clark, and Hayes]{sigopt-web-page}
Clark,~S.; Hayes,~P. {SigOpt} {W}eb page. \url{https://sigopt.com}, 2019;
  \url{https://sigopt.com}\relax
\mciteBstWouldAddEndPuncttrue
\mciteSetBstMidEndSepPunct{\mcitedefaultmidpunct}
{\mcitedefaultendpunct}{\mcitedefaultseppunct}\relax
\EndOfBibitem
\bibitem[Chase(1998)]{JANAF}
Chase,~M. NIST-JANAF Thermochemical Tables. \emph{American Institute of
  Physics} \textbf{1998}, \emph{4}\relax
\mciteBstWouldAddEndPuncttrue
\mciteSetBstMidEndSepPunct{\mcitedefaultmidpunct}
{\mcitedefaultendpunct}{\mcitedefaultseppunct}\relax
\EndOfBibitem
\end{mcitethebibliography}

\section{Supplementary Information}

\subsection{Model details}

The per-site model architecture presented here is an adaptation of the CGCNN developed by \cite{Xie2018}. The model takes three-dimensional crystal structures as input (Fig. \ref{architecture}a) and represents them as undirected multigraphs. The graph representation encodes atoms as nodes and connectivity as edges. Each site in the crystal is designated by a feature vector encoding its atomic properties and a neighbor list of sites within a cutoff radius. The neighbor connections, graph edges, are featurized with the distance between atoms. Distances between sites are computed in a periodicity-aware manner and, as such, the crystal graph can have multiple edges between a given pair of nodes. The crystal graph is passed through convolutional layers, which iteratively update the nodal feature vectors with information from their associated neighbors and edge connectivity, thus automatically learning crystal site representations informed by their unique chemical environments (Fig. \ref{architecture}b). The model then passes the convolved per-site representations through several fully connected hidden layers, and finally to an output layer to yield a scalar target property prediction at each site (Fig. \ref{architecture}c). The model is trained by minimizing the difference, defined by a loss function $L(\mathbf{p},\mathbf{\tilde{p}})$, between target (DFT-calculated) and predicted per-site property values, $\mathbf{p}$ and $\mathbf{\tilde{p}}$, respectively. The vector of per-site predictions is computed as $\mathbf{\tilde{p}} = f(\mathcal{C}; \mathbf{W})$, where $f$ is the CGCNN network, $\mathcal{C}$ is the input crystal, and $\mathbf{W}$ is the set of weights that parametrizes the model layers \citep{Xie2018}. During training, model weights $\mathbf{W}$ are optimized to minimize $L(\mathbf{p},\mathbf{\tilde{p}}=f(\mathcal{C}; \mathbf{W}))$ through iterative updates calculated by backpropagation and stochastic gradient descent. We additionally test an analogous per-site model adapted from PAINN\cite{Schutt2021}.

\begin{table}
  \caption{Fitted bond-valence sum parameters, R$_0$ and B, for O-M bonds.} 
  \label{bond_valence_parameters}
  \centering
  \begin{adjustbox}{width=0.8\textwidth}
  \begin{tabular}{lcccc}
    \toprule
    M & \shortstack{O 2p-band center \\ R$_0$} &  \shortstack{O 2p-band center \\ B} & \shortstack{O Bader charge \\ R$_0$} & \shortstack{O Bader charge \\ B} \\
    \midrule
    Ti & 2.12 & 0.37 & 1.66 & 0.40 \\
    V & 2.15 & 0.41 & 1.48 & 0.56 \\
    Cr & 2.10 & 0.41 & 1.32 & 0.84 \\
    Mn & 2.08 & 0.40 & 1.40 & 0.74 \\
    Fe & 2.05 & 0.40 & 1.68 & 0.40 \\
    Co & 2.02 & 0.46 & 0.46 & 1.69 \\
    Ni & 1.96 & 0.27 & 1.55 & 0.48 \\
    Zr & 2.29 & 0.51 & 1.89 & 0.36 \\
    Nb & 2.52 & 0.63 & 1.67 & 0.56 \\
    Mo & 2.35 & 0.53 & 1.55 & 0.72 \\
    Ru & 2.29 & 0.52 & 1.15 & 1.13 \\
    Rh & 2.25 & 0.62 & 1.32 & 0.75 \\
    Pd & 2.16 & 0.74 & 1.49 & 0.56 \\
    Hf & 2.31 & 0.47 & 1.84 & 0.41 \\
    Ta & 2.60 & 1.03 & 1.71 & 0.56 \\
    Re & 2.64 & 0.76 & 1.55 & 0.64 \\
    Os & 2.37 & 0.53 & 1.04 & 1.45 \\
    Ir & 2.44 & 0.69 & 1.00 & 1.16 \\
    Pt & 2.27 & 0.57 & 1.28 & 0.73 \\
    
    \bottomrule
  \end{tabular}
  \end{adjustbox}
\end{table}

\end{document}